\documentclass[aps]{revtex4-1}
\usepackage{epsfig}
\usepackage{graphicx}
\usepackage{epstopdf}
\usepackage{amsmath}
\usepackage{verbatim}

\newcommand{\beq}{\begin{eqnarray}}
\newcommand{\eeq}{\end{eqnarray}}
\usepackage{amssymb}
\usepackage{hyperref}
\hypersetup{colorlinks}
\begin{document}
\title{Electronic Energy Transfer: Localized Operator Partitioning of Electronic Energy in Composite Quantum Systems}
\author{Yaser Khan}
\author{Paul Brumer}
\affiliation{Chemical Physics Theory Group, \\ Department of Chemistry and \\
Center for Quantum Information and Quantum Control\\
University of Toronto, Toronto, M5S 3H6, Canada}
\date{\today}
\begin{abstract}
A Hamiltonian based approach using
spatially localized projection operators is introduced to give
precise meaning to the
chemically intuitive idea of the electronic energy on a quantum subsystem.
This definition facilitates the study of electronic energy transfer
in arbitrarily coupled quantum systems. In particular,
the decomposition scheme can be applied
to molecular components that are strongly interacting
(with significant orbital overlap)
as well as to isolated fragments.  The result leads to the proper
electronic energy at all internuclear distances, including the case of
separated  fragments,
and reduces to the well-known
F\"orster and Dexter results in their respective limits.
Numerical calculations of coherent energy and charge transfer
dynamics in simple model systems are presented and the  effect of collisionally induced decoherence is
examined.
\end{abstract}
\maketitle

\section{Introduction}
Measuring, predicting and
ultimately controlling the rate of electronic energy transfer (EET) within and between molecules
is currently of great scientific and technological interest.
For example, technologically understanding  natural EET processes \cite{scholesrev,sundstrom,grondelle}
may well contribute to of our ability to design efficient photovoltaic devices
\cite{pv1,pv2,pv3,pv4,pv5}.

Ideally, given a large molecular system or molecular network through which
electronic energy flows, we would proceed as follows: After obtaining the
system Hamiltonian $H$, we would solve the time dependent Schr\"{o}dinger equation
to obtain the wavefunction $\Psi(t)$ for a given initial wavefunction $\Psi(0)$. To track the electronic energy $E_A$ on,
e.g., site A, we would compute $E_A = \langle \Psi(t)|\mathcal H_A |\Psi(t)\rangle$,
where $\mathcal H_A$ is the operator which corresponds to electronic energy on site $A$.
The computed $E_A$, being an integral over an operator on the electronic
degrees of freedom, would automoatically include decoherence effects
arising from other degrees of freedom\cite{ignacio}.

Clearly this is computationally difficult for realistic molecular systems.
Nonetheless, one expects, from basic quantum principles, that there exists such a real valued operator $\mathcal H_A$. For this reason it is surprising
that we find no study of such an operator for electronic site energies.
Rather, EET is most often studied via a series of approximate schemes
reliant upon simplifying models that are applicable in restricted circumstances.

In an effort to better quantify the rates of EET, the role of vibrational
modes in decohering EET, the effects of assorted coupling scenarios in
molecular systems, the role of orbital overlap between sites, etc. we
obtain below a meaningful $\mathcal H_A$ operator that
describes electronic energy on a site. The operator is valid at all
internuclear distances and, as such, resolves
problems (elucidated below) arising from electron antisymmetrization at all distances.
In addition, the model is applicable in all electronic energy transfer regimes, i.e. weak,
intermediate and strong electronic coupling, and reduces to
 known results, e.g. the F\"orster and Dexter results, in the appropriate limit.

This paper is organized as follows: Section~\ref{sec:motivation} motivates
the problem of defining electronic energy on a site via the simple example of the
Hydrogen molecule, $H_2$. Section~\ref{sec:definition} then introduces the Hamiltonian
decomposition that effectively resolves this issue.   Limiting cases are discussed
in Section~\ref{sec:application}. Numerical results of coherent energy and charge transfer
dynamics between Hydrogen atoms and molecules are presented in
Section~\ref{sec:compute} and a summary with
conclusion in Section~\ref{sec:conclusion}.

\section{Motivation} \label{sec:motivation}

A central issue associated with defining an $\mathcal{H}_A$ arises due to electron antisymmetrization.  For simplicity, consider a molecule composed of two atomic fragments
$A$ and $B$, where interest is in the electronic energy in Site A or B.
Motivation arises from considering even the simplest of cases, i.e., hydrogen.  Here, the Hamiltonian for HD (chosen to simplify the issue by using distinguishable nuclei) is of the form
\begin{equation}
H=H_{A1}+H_{B2}+V_{1,2} \label{eq1}
\end{equation}
where the $A$ and $B$ label the two nuclei, and $1$ and $2$ label the electrons.
The electronic Hamiltonian is (with $\hbar=1$)
\begin{equation}
  H=
    \underbrace{- \frac{1}{2}\nabla_1^2-\frac{1}{r_{A1}}}_{H_{A1}}
    \underbrace{- \frac{1}{2}\nabla_2^2-\frac{1}{r_{B2}}}_{H_{B2}}
  + \underbrace{\frac{1}{R}+\frac{1}{r_{12}}-\frac{1}{r_{A2}}-\frac{1}{r_{B1}}}_{V_{1,2}}~.\label{eq2}
\end{equation}
$R$ is the internuclear distance, $r_{\alpha i}$ is the distance of electron $i$ from nucleus $\alpha$, and $\nabla_i^2$ is the Laplacian associated with   the i$^{th}$ electron.
The exact wavefunction, $\vert\Psi_{1,2}\rangle$,
satisfies
\begin{equation}
H \vert\Psi_{1,2}\rangle = E \vert\Psi_{1,2}\rangle .
\end{equation}

We highlight the definition problem by calculating the electronic energy $E_A$ of subsystem
$A$ for an initially na\"ive  choice of the subsystem
Hamiltonian at large $A-B$ interatomic separation.  Chemical intuition requires that the resultant $E_A$
be equal to the isolated subsystem energy
for any acceptable definition of $\mathcal H_A$. We calculate this energy first by incorrectly assuming that the
electrons in the system are distinguishable and then contrast the result
with the calculation that takes proper account of electron indistinguishability
through wavefunction antisymmetrization.

\subsection{Ignoring Electron Indistinguishability}
\label{sec:dist_elec}
In the absence of antisymmetrization, the molecular wavefunction
is, to a first approximation, a simple Hartree product
\begin{equation}
\vert\Psi_{1,2}\rangle \equiv \vert A_1 B_2 \rangle = \vert A_1 \rangle \vert B_2 \rangle .
\end{equation}
where $H_{A1} \vert A_1 \rangle = E_A \vert A_1 \rangle$ and
$H_{B2} \vert B_2 \rangle = E_B \vert B_2 \rangle$, and
$\vert A_1 \rangle$ and $\vert B_2 \rangle$ are normalized
spin orbitals centered on the respective atoms.

The energy of the molecule $AB$ is then
\begin{equation}
  E= \langle \Psi_{1,2}\vert H \vert\Psi_{1,2}\rangle
				= E_A + E_B + V_{AB}~,
\end{equation}
where
 \begin{eqnarray}
  E_A &=& \langle A_1 \vert H_{A1} \vert A_1 \rangle
=  \langle A_1 \vert - \frac{1}{2}\nabla_1^2 \vert A_1 \rangle +  \langle A_1 \vert -\frac{1}{r_{A1}} \vert A_1 \rangle
\equiv T_A + V_A  \nonumber \\
  E_B &=& \langle B_2 \vert H_{B2} \vert B_2 \rangle
= \langle B_2 \vert - \frac{1}{2}\nabla_2^2 \vert B_2 \rangle + \langle B_2 \vert -\frac{1}{r_{B2}} \vert B_2 \rangle
\equiv T_B + V_B  \nonumber \\
  V_{AB} &=& \left \langle A_1B_2 \left \vert  {\frac{1}{R_{\phantom{1}}}}
                                              +{\frac{1}{r_{12}} }
                                              -{\frac{1}{r_{A2}} }
                                              -{\frac{1}{r_{B1}} }
                     \right \vert A_1B_2 \right \rangle
\label{eq6} \end{eqnarray}
 As $R \to \infty$, $V_{AB} \to 0$ since all terms are $O(1/R)$.  The results meet  expectations: at infinite separation the
energy associated with  atoms $A$ or $B$ is equal to the energy of the
respective isolated atoms.  Hence it appears that $\mathcal{H}_{A1} = H_{A1}$ is a possible candidate for measuring  electronic energy on site A.

\subsection{Antisymmetrized Electrons}
\label{sec:ident_elec}
Consider now the case where  proper
account is taken of wavefunction antisymmetrization. Here, the wavefunction is
\begin{equation}
  \vert \Psi_{1,2}^S \rangle \equiv \frac{1}{\sqrt{2}} \vert A_1B_2 - B_1A_2 \rangle
\end{equation}
where the subscripts denote the electron label on the indicated atom and the superscript ``S" denotes properly antisymmetrized quantities.

Once again, we na\"ively define [Eq. (\ref{eq2})]
the energy on atoms $A$ and $B$,via H$_{A1}$ and H$_{B2}$, and the interaction energy $V_{AB}^S$,
as:
\begin{eqnarray}
    & E_A^S &= \langle \Psi_{1,2}^S \vert H_{A1} \vert \Psi_{1,2}^S \rangle \nonumber \\
    & E_B^S &= \langle \Psi_{1,2}^S \vert H_{B2} \vert \Psi_{1,2}^S \rangle  \nonumber \\
    & V_{AB}^S &= \langle \Psi_{1,2}^S\vert V_{1,2} \vert \Psi_{1,2}^S \rangle .
  \end{eqnarray}
Expanding $E_A^S$ gives
 \begin{eqnarray}
  E_A^S &=& \frac{1}{2} [ \langle A_1B_2\vert H_{A1} \vert A_1B_2 \rangle
      - \langle A_1B_2\vert H_{A1} \vert B_1A_2 \rangle \nonumber \\
     & -& \langle B_1A_2\vert H_{A1} \vert A_1B_2 \rangle
      + \langle B_1A_2\vert H_{A1} \vert B_1A_2 \rangle ]~.
\end{eqnarray}
which, in the large $R$ limit,  becomes
\begin{equation}
  E_A^S = \frac{1}{2}(E_A + T_B)
        = \frac{1}{2}(T_A + T_B + V_A)~.
\end{equation}
Similarly,

\begin{equation}
  E_B^S \equiv  \frac{1}{2}  \langle A_1B_2 - B_1A_2 \vert H_{B2} \vert A_1B_2 - B_1A_2 \rangle
        =       \frac{1}{2}(T_A + T_B + V_B)~.
\end{equation}
Finally, consider the interaction term
\begin{equation}
  V_{AB}^S = \frac{1}{2}  \langle A_1B_2 - B_1A_2 \vert V_{1,2} \vert A_1B_2 - B_1A_2 \rangle~.
\end{equation}
As $R \rightarrow \infty$,  $V_{AB}^S = V_A + V_B$, where
\begin{equation}
     \langle B_1A_2 \vert V_{1,2} \vert B_1A_2 \rangle
           = \displaystyle{\int(B_1^{*}B_1) \left (
                 {\frac{1}{r_{12}}-\frac{1}{r_{A2}}-\frac{1}{r_{B1}}}
               \right ) (A_2^{*}A_2) \, \mathrm{d} \tau_{1,2}} \\
           \equiv  V_{A} + V_{B} \quad \hbox{as} \quad R \to \infty
 \end{equation}
Hence, as $R \rightarrow \infty$
\begin{eqnarray}
  V_{AB}^S &=& \frac{1}{2}(V_A + V_B) \nonumber \\
  E^S_A & = & \frac{1}{2}(T_A + T_B + V_A) \nonumber  \\
  E_B^S & = & \frac{1}{2}(T_A + T_B + V_B)
\label{eq14} \end{eqnarray}
Thus, for the case of the antisymmetrized wavefunctions, although the total energy
\begin{equation}
  E^S = E_A^S + E_B^S + V_{AB}^S
      = E_A + E_B .
\end{equation}
 is correct, the energy partitioning amongst individual site energies $E_A^S$ and $E_B^S$ [Eq. (\ref{eq14})] is wrong.

This issue is far from being new. For example, Margenau raised the same  concern in 
considering antisymmetrization issues \cite{margenau}  and attributed its resolution to a
sophisticated measurement problem. Alternative suggestions
would be, for example, that
antisymmetrization becomes unnecessary at large distances between
electrons. This,
however, is not the case. That is,
requirements for electron exchange do not arise via a force
that diminishes with $R$. Rather, examination of the proposed
site operator $\mathcal{H}_i$ shows, as discussed below,
that    the    likely    problem    is    that
$H_{A1},H_{B2}$  themselves unjustifiably  distinguish
electron 1 from electron 2.

\section{Defining the Subsystem Hamiltonian}
\label{sec:definition}
Consider now the general case. The electronic Hamiltonian of the composite system, $H$, may be
written explicitly in terms of its one and two electron terms,
$h$ and $g$, as
 \begin{equation} H 	=	\sum_i^N h(x_i) + \frac{1}{2} \sum_{i,j}^{N}\!^{'}g(x_i,x_j) \label{Hamil}
\end{equation}
where $h(x_i)=- \frac{1}{2}\nabla_i^2 - \sum_{\beta} {Z_{\beta}}/{r_{\beta i}}$,
$g(x_i,x_j)= {1}/{r_{ij}}$, $i$ and $j$ label the electrons, $\alpha$ labels the nuclei
and $Z_{\beta}$ is the charge, in atomic units, on  the $\beta$ nucleus. The
prime on the second summation denotes $i\ne j$.

Since the electrons are indistinguishable and all interactions are
two-body we can simplify the integrand when the Hamiltonian appears
inside an integral over electronic coordinates as follows:

\begin{equation} \begin{split}
    \langle \Psi(x_1,x_2,\ldots,x_N) &\vert H \vert \Psi(x_1,x_2,\ldots,x_N) \rangle \\
    &= \langle \Psi(x_1,x_2,\ldots,x_N) \vert
	\displaystyle{\sum_i^N h(x_i) + \frac{1}{2} \sum_{i,j}^{N}\!^{'}g(x_i,x_j)}
        \vert \Psi(x_1,x_2,\ldots,x_N) \rangle \\
    &= \langle \Psi(x_1,x_2,\ldots,x_N) \vert
	\displaystyle{N h(x_1) + \frac{N(N-1)}{2} g(x_1,x_2)}
        \vert \Psi(x_1,x_2,\ldots,x_N) \rangle \\
    &= \langle \Psi(x_1,x_2,\ldots,x_N) \vert \mathcal{H} \vert \Psi(x_1,x_2,\ldots,x_N) \rangle   \label{Haverage} \end{split}
\end{equation}
where $\Psi(x_1,x_2,\ldots,x_N)$ is a properly antisymmetrized wavefunction, and
\begin{equation}
  \mathcal{H} = N h(x_1) + \frac{N(N-1)}{2} g(x_1,x_2)~.\label{reducedH}
\end{equation}
Giving Eq. (\ref{Haverage}) we can use Eq. (\ref{reducedH}) in place of the Hamiltonian whenever it appears inside an integral over all electronic coordinates.
This substitution is also valid if the integrand is of the form $H\Omega$ for any  $\Omega$ that is
symmetric with respect to pairwise electron interchange and is composed of one and two electron
operators. This is used below    to simplify the notation.

Without loss of generality we assume that our interest is in the electronic energy on a single subsystem,
denoted $A$. The subsystems, each labeled by the index $\alpha$, are defined
spatially such that  they collectively span all space. Given the system Hamiltonian
$H$, we set out to define a subsystem Hamiltonian, $\mathcal{H}_A$, such that
the electronic energy of the subsystem, $E_{A}$, is given by
$E_A={\rm Tr}[\hat{\rho} \mathcal{H}_A]$, where $\hat{\rho} = |\Psi\rangle \langle\Psi|$ is
 the system density operator. In coordinate space
$\Psi(x_1,x_2,\ldots,x_N)$ is the normalized antisymmetric wavefunction
 of the system, and the trace is taken over all electronic coordinates.
Specifically, we look for a form $\mathcal{H}_A = \Gamma_A \mathcal{H}$, where $\Gamma_A$ is a superoperator.
Interestingly, the choice of $\Gamma_A$ depends upon whether interest
is in stationary or non-stationary states.

Three principles guide the choice of $\mathcal{H}_A$ and associated fragment energy $E_A = \textrm{Tr}[\hat{\rho}\mathcal{H}_A]$:

(1) The fragment energy $E_A$ must be real, for both stationary $\hat{\rho}$ as well as time dependent $\hat{\rho}(t)$.

(2) The energy $E_A$ must reduce to the correct energy for an independent fragment A.

(3)  The operator $\mathcal{H}_A$ must be symmetric with respect to electron interchange.

The first two requirements are evident, whereas the third benefits from some justification.  Specifically, consider
\begin{equation} E_A = \textrm{Tr}[\hat{\rho}\mathcal{H}_A] = \langle \Psi(t)|\mathcal{H}_A| \Psi(t)\rangle \end{equation}
Since $\hat{\rho}$ is symmetric with respect to electron interchange, if $\mathcal{H}_A$ were antisymmetric with respect to electron interchange it would result in an $E_A$ which is similarly antisymmetric.  Such an energy, whose sign would depend on electron identity, is non-physical. As a consequence, $\mathcal{H}_A$ must be symmetric.

In addition, the method should reproduce standard results (e.g., F\"{o}rster and Dexter energy transfer) when applied to those cases.  All of these characteristics are verified for the $\mathcal{H}_A$ definition introduced below.

\emph{Stationary state}.  The choice of $\Gamma_A$ in the definition of the subsystem Hamiltonian
 $\mathcal H_A$ is motivated by physical intuition in conjunction with the requirements above.
In particular,  the proposed   expression for the subsystem Hamiltonian is
\begin{equation} \begin{split}
\mathcal  H_A &= \Gamma_A\mathcal H = \Gamma_{A}[N h(x_1)
			+ \frac{N(N-1)}{2} g(x_1,x_2)] \\
		&\equiv        N\Theta_{A}(x_1) h(x_1)
			+ \frac{N(N-1)}{2}\sum_\alpha \Theta_{A,\alpha}(x_1,x_2)g(x_1,x_2)
		 \end{split}
\end{equation}
where
\begin{align}
  \Theta_A(x_1) = \begin{cases}
	1 & \text{if~} x_1 \,\epsilon \, A \\
	0 & \text{otherwise}~.
  \end{cases}
  \label{proj1}
\end{align}
Here $\Theta_{A,\alpha}(x_1,x_2)$ is a symmetric spatial projection operator
\begin{align}
  \Theta_{A,\alpha}(x_1,x_2)	&=	 \frac{1}{2}(\Theta_A(x_1)\Theta_\alpha(x_2)+\Theta_\alpha(x_1)\Theta_A(x_2))
  \label{proj2}
\end{align}
such that
\begin{align}
  \Theta_{\alpha,A}(x_1,x_2)	&=	 \Theta_{A,\alpha}(x_1,x_2) \nonumber \\
  \Theta_{A,\alpha}(x_1,x_2)	&=	 \Theta_{\alpha,A}(x_1,x_2) .
\end{align}
Note the completeness of the projection operators:
\begin{equation}
      \mathbf{1}(x_1) = \displaystyle{\sum_{\alpha}\Theta_{\alpha}(x_1)}
	\label{ident1}
\end{equation}
and that Li and Parr \cite{li} have shown that  $\Theta_{A,B}(x_i,x_j)$
 is ``physically sound'' \cite{sound} 
and that it preserves the completeness relation
which generalizes, for two electrons, to
\begin{equation}
  \mathbf{1}(x_1,x_2) 	= 	\sum_{\alpha,\beta} \Theta_{\alpha,\beta}(x_1,x_2).
  \label{ident2}
\end{equation}
Here  $\alpha = A, B$ and $\beta = A, B$.  With these definitions we can define
$\mathcal H_A$ concisely as
\begin{equation}
\mathcal H_A 	=	Nh_A(x_1) + \frac{N(N-1)}{2}g_{A,A}(x_1,x_2)
			+ \frac{N(N-1)}{2}\sum_{\alpha \ne A} g_{A,\alpha}(x_1,x_2)
\label{eqn:deftiha}
\end{equation}
where
$h_A(x_1) = \Theta_A(x_1) h(x_1)$,
$g_{A,A}(x_1,x_2) = \Theta_{A,A}(x_1,x_2)g(x_1,x_2)$ and
$g_{A,\alpha}(x_1,x_2) =  \Theta_{A,\alpha}(x_1,x_2)g(x_1,x_2)$.  Hence
$h_A(x_1)$ is the one electron energy of the  electron density that
resides inside region $A$, $g_{A,\alpha}(x_1,x_2)$ is the two electron energy when
only one of the (indistinguishable) electrons is in region $A$, and $g_{A,A}(x_1,x_2)$
is the two electron energy with both electrons  in region $A$.

The resultant expression for the subsystem electronic energy, $E_A$, is
valid for general time independent densities, $\rho = \sum_i p_i|\Psi_i\rangle\langle\Psi_i|$, giving the electronic energy on site A as

\begin{equation}
  E_A 	= {\rm Tr}\{\mathcal{H}_A\hat{\rho}\}
 	= \sum_i p_i {\rm Tr} \{\mathcal H_A |\Psi_i\rangle\langle\Psi_i| \}
	= {\rm Tr_{12}} \{ \mathcal H_A \hat{\rho}_{12} \}
\end{equation}
where $\hat{\rho}_{12}=\sum_i p_i \int dx_3dx_4\ldots dx_N
\langle {\bf x}|\Psi_i\rangle\langle\Psi_i| {\bf x} \rangle$, and ${\bf x}$
denotes $(x_1,x_2, \ldots, x_N)$.
The coordinate representation of $\hat{\rho}_{12}$ is
$\langle x_1^{'},x_2^{'}\vert\hat{\rho}_{12} \vert x_1,x_2\rangle$.
In evaluating  terms like  $ {\rm Tr_{12}} \{ h_A(x_1) \hat{\rho}_{12} \}$,
 we calculate
$ {\rm Tr_{12}} \{ h_A(x_1) \hat{\rho}_{12} \}
= {\int}_{x_1^{'}=x_1 \atop {x_2^{'}=x_2}}  dx_1dx_2 h_A(x_1)
\langle x_1^{'},x_2^{'}\vert\hat{\rho}_{12} \vert x_1,x_2\rangle$,
where we put $x_1^{'}=x_1$ and $x_2^{'}=x_2$ after operating
with $h_A(x_1)$ but before carrying out the integration  \cite{mcweeny}.
That $E_A$ is real follows from the reality of the terms in the sum
of products above. All the terms in the subsystem Hamiltonian $\mathcal H_A$
(Eq.~\ref{eqn:deftiha}) are evidently real and, for a system in a stationary state,
the coordinate representation of the density matrix is also  real.

\emph{Time Evolving State}.  For a system in a non-stationary state, significant for energy transfer
studies, the wavefunction is described by a
superposition of energy eigenstates. The matrix elements of the time dependent
density operator, $\hat{\rho}(t) = \sum_i p_i|{\Psi}_i(t)\rangle\langle{\Psi}_i(t)|$, in the coordinate representation are generally complex. We therefore
need to generalize  $\Gamma_A$ to ensure
that $E_A(t)$ is real.  To examine the issue note that
given a one particle complete orthonormal basis $\{\chi_k\}$ one can
express the subsystem energy $E_A(t)$ as
\begin{equation}
            E_A(t)				= {\rm Tr}\{\mathcal{H}_A\hat{\rho}(t)\}
					= {\rm Tr_{12}}\{\mathcal{H}_A\hat{\rho_{12}}(t)\}
					= \sum_{jklm}  \rho_{lm,jk}\mathcal H^{jk,lm}_{A} \label{EA}
\end{equation}
where
\begin{equation}
	\rho_{jk,lm}	= \langle \chi_j(1)\chi_k(2) \vert \hat{\rho}_{12} \vert \chi_l(1)\chi_m(2)\rangle,
\end{equation}
and
\begin{eqnarray}
	\mathcal H^{jk,lm}_{A} &	=& \langle \chi_j(1)\chi_k(2) \vert \Gamma_A \mathcal H \vert \chi_l(1)\chi_m(2)\rangle  \nonumber \\
				&=& N\langle \chi_j(1) \vert \Gamma_A h(1) \vert \chi_l(1) \rangle \delta_{km}
					+ \frac{N(N-1)}{2}\langle \chi_j(1)\chi_k(2) \vert \Gamma_A g(1,2) \vert \chi_l(1)\chi_m(2)\rangle~.\label{Hbasis}
\end{eqnarray}
Consider each term in  Eq. (\ref{EA}).  Since
$\hat{\rho}(t)$ is Hermitian, $\rho_{jk,lm}^{*}(t) = \rho_{lm,jk}(t)$.
While each term in the two electron contribution
is generally complex, the fact that the projection operator, $\Gamma_A$, and the
two electron interaction operator, $g(x_1,x_2) =1/{x_{12}}= 1/|x_1 - x_2|$,
are real ensures that the terms in the expression for $E_A(t)$ sum
in pairs to give a real result. In particular this is because
$\langle \chi_j(1)\chi_k(2) \vert \Gamma_A g(x_1,x_2) \vert \chi_l(1)\chi_m(2) \rangle^{*}
=\langle \chi_j(1)\chi_k(2) \vert \Gamma_A g(x_1,x_2) \vert \chi_l(1)\chi_m(2) \rangle$
for any choice of region $A$ subject to the conditions on the projection operators
discussed above (Eq.~\ref{proj1}-\ref{ident2}).

By contrast, the one electron term
$\langle \chi_j(1) \vert \Gamma_A h(x_1) \vert \chi_l(1) \rangle$ contains a kinetic energy
contribution  $T = -\nabla_1^2/2$,
which is  Hermitian only for particular
partitions of space such that the wavefunction satisfies particular boundary
conditions.   These boundary conditions are trivially satisfied when the wavefunction
and its first derivatives vanish at the boundary. This is the case, for example,
when we consider the entire system, where its boundaries are at infinity.
The wavefunctions also vanish in the region between two subsystems
if they are sufficiently separated. This is, however, generally not the case.

To bypass this difficulty we recognize the possibility of utilizing non-Hermitian operators that have real eigenvalues  \cite{nonhermitian}. We introduce  an alternative Hamiltonian-based
real space partitioning approach that is
computationally efficient and applicable to arbitrary
electronic states of the system. Specifically, we average the sum of the matrix with its transpose.
The generalized subsystem Hamiltonian, $\mathcal H_A$,
obtained in this way is applicable to both time dependent and time
independent densities and is of the form:
\begin{equation}
  \mathcal H_A	= \Gamma_A \mathcal H
		=	N\frac{\Theta_{A}(x_1)h(x_1)+ h(x_1^{'})\Theta_{A}(x_1^{'})}{2}
				+ \frac{N(N-1)}{2}\sum_\alpha \Theta_{A,\alpha}(x_1,x_2)g(x_1,x_2)
\end{equation}

\begin{equation}
  \mathcal H_A	=	\frac{N}{2}\left (h_A(x_1)+ h_A(x_1^{'})\right ) + \frac{N(N-1)}{2}g_{A,A}(x_1,x_2)
				+ \frac{N(N-1)}{2}\sum_{\alpha \ne A} g_{A,\alpha}(x_1,x_2)~ . \label{eqn:deftdha}
\end{equation}
The expectation value of energy, ${\rm Tr}\{\rho \mathcal H_A\}$,
using Eq.~(\ref{eqn:deftdha})  for a time-independent density, reduces to the
time-independent result [Eq.~\ref{eqn:deftiha}].

Note that  $\mathcal H_A$  is a
non-Hermitian operator with real eigenvalues. That the eigenvalues of the
operator are real is demonstrated by the fact that the expectation value of
$\mathcal H_A$ is real for any state of the system. The non-Hermitian
character of the $\mathcal H_A$ results from the fact that the subsystem,
as an open system, evolves nonunitarily in time \cite{nonhermitian}.

\section{Application to Energy Transfer}
\label{sec:application}
The above results are applicable to molecular systems of any constituency, and with components at any intermolecular distance.  Here  we apply the  definition of the subsystem Hamiltonian
to donor-acceptor systems and illustrate the agreement between results calculated
based on $\mathcal{H}_A$ and well known limits that are applicable when the subsystems are suitably separated.

Energy Transfer problems are solved in
different limiting regimes that are classified
according to the spatial proximity of, and coupling strength between,
interacting molecules. We will follow the development of these limits
as outlined by Parson \cite{parson}. To illustrate analytically the agreement of our result with the F\"orster and Dexter limits,
 consider a sample calculation in the limit of weak coupling and insignificant subsystem overlap.
The composite system is partitioned  into two distinct parts, $A$ and $B$  with normalized subsystem eigenstates
 $\vert a_i\rangle$ and $\vert b_j\rangle$. To simplify the algebra we assume that $A$ and $B$
are far enough apart that the overlap, $\langle a_i\vert b_j\rangle$, is sufficiently
small so as not to require renormalization of the product wavefunction without necessarily
being zero.
We denote     Hamiltonians of the two isolated
systems by  $H_A$ and $H_B$; by definition then
$\langle a_i(1) \vert H_A(1) \vert a_j(1) \rangle = E_{a_i}\delta_{ij}$, and
similarly for $B$.

\subsection{Two-electron Case}

For $A$ and $B$ sufficiently separated the system Hamiltonian of the non-interacting composite
system  is given by $H = H_A + H_B$ and  the eigenstates of the
composite system are
$\vert \psi_{ij} \rangle = \frac{1}{\sqrt{2}}\vert a_i(1)b_j(2)-b_j(1)a_i(2) \rangle$.
A general state, $\Psi$, of the system is then a superposition
$\vert \Psi \rangle = \sum_{ij} c_{ij} \vert \psi_{ij} \rangle$,  with the  density matrix
 given by $\hat{\rho} = \vert \Psi \rangle \langle \Psi \vert$. Using the subsystem Hamiltonian
$\mathcal H_A$ the energy of subsystem $A$, $E_A$, in state $\Psi$ is
\begin{eqnarray}
    E_A 	&=&	
		\langle \Psi \vert   \mathcal H_A \vert \Psi \rangle \nonumber \\
	&=	&\left \langle \sum_{ij} c_{ij}\psi_{ij} \right \vert  \mathcal H_A \left \vert \sum_{kl} c_{kl}\psi_{kl} \right \rangle
	=	\sum_{ijkl} c_{ij}^{*}c_{kl}\langle \psi_{ij} \vert  \mathcal H_A \vert \psi_{kl} \rangle
\end{eqnarray}
In the limit of well separated subsystems the wavefunction of the system
vanishes at the boundary between the subsystems. As discussed in the preceding
section on time evolving states  we are therefore
justified in using $\mathcal H_A$ as given in Eq.~(\ref{eqn:deftiha}). Focusing on
the $(ij, kl)$ matrix element gives
\begin{equation}
	\langle \psi_{ij} \vert  \mathcal H_A \vert \psi_{kl} \rangle
		= N\langle \psi_{ij} \vert \Theta_A(1) h(1)  \vert \psi_{kl} \rangle + 	
 	       \frac{N(N-1)}{2} \langle \psi_{ij} \vert \sum_\alpha\Theta_{A,\alpha}(1,2) g(1,2)  \vert \psi_{kl} \rangle  	
\end{equation}

Moreover
\begin{eqnarray}
	\langle \psi_{ij} \vert h_A(1) &= &	\langle \psi_{ij} \vert \Theta_A(1)h(1)
	= \frac{1}{\sqrt{2}} \langle a_i(1) b_j(2) - b_j(1) a_i(2) \vert \Theta_A(1)h(1) \nonumber \\
		&=&  \frac{1}{\sqrt{2}} \langle a_i(1) b_j(2) \vert h(1)
\end{eqnarray}
and
\begin{equation}
	\langle \psi_{ij} \vert g_{A,A}(1,2) =	\langle \psi_{ij} \vert \Theta_{A,A}(1,2)g(1,2)
	= \frac{1}{\sqrt{2}} \langle a_i(1) b_j(2) - b_j(1) a_i(2) \vert \Theta_A(1)\Theta_A(2) g(1,2) = 0  
\end{equation}

\begin{eqnarray}
	\langle \psi_{ij} \vert g_{A,B}(1,2)  &=& \langle \psi_{ij} \vert \Theta_{A,B}(1,2)g(1,2) \nonumber \\
	&=& \frac{1}{\sqrt{2}} \langle a_i(1) b_j(2) - b_j(1) a_i(2) \vert
	       \frac{1}{2}(\Theta_A(1)\Theta_B(2)+\Theta_B(1)\Theta_A(2))g(1,2)  \nonumber \\
		&=& \frac{1}{2\sqrt{2}} \langle a_i(1) b_j(2) - b_j(1) a_i(2) \vert g(1,2)
\end{eqnarray}
Hence,
\begin{eqnarray}
	\langle \psi_{ij} \vert \mathcal{ H}_A \vert \psi_{kl} \rangle &= & \frac{1}{2} \langle a_i(1) b_j(2) - b_j(1) a_i(2) \vert \Theta_A(1) \mathcal H \vert  a_k(1) b_l(2) - b_l(1) a_k(2)                      \rangle \nonumber \\
	&=& \frac{N}{2} \langle a_i(1) b_j(2) - b_j(1) a_i(2) \vert \Theta_A(1) h(1) \vert  a_k(1) b_l(2) - b_l(1) a_k(2) \rangle \nonumber \\
	    && +~~ \frac{N(N-1)}{2} \sum_\alpha \langle a_i(1) b_j(2) \nonumber \\ &&-~~ b_j(1) a_i(2) \vert \Theta_{A,B}(1,2) g(1,2) \vert  a_k(1) b_l(2) - b_l(1) a_k(2)  \rangle~.
\end{eqnarray}
With the number of electrons N = 2, and the summation over partitions $A$ and $B$,
\begin{eqnarray}
	\langle\psi_{ij}|\mathcal{H}_A|\psi_{kl}\rangle &=& \langle a_i(1) b_j(2) \vert h(1) \vert  a_k(1) b_l(2) - b_l(1) a_k(2) \rangle \nonumber \\
	    && +~~ \frac{1}{2} \sum_\alpha \frac{1}{2} \langle a_i(1) b_j(2) - b_j(1) a_i(2) \vert g(1,2) \vert  a_k(1) b_l(2) - b_l(1) a_k(2) \rangle \nonumber \\
	&=& \underbrace{\langle a_i \vert h(1) \vert  a_k  \rangle }_{E_{A_i}\delta_{ik}}
	  + {\tfrac{1}{2}}\underbrace{\langle a_i b_j \vert g(1,2) \vert  a_k b_l \rangle}_{J_{ijkl}}
	  - {\tfrac{1}{2}}\underbrace{\langle a_i b_j \vert g(1,2) \vert  b_l a_k \rangle}_{K_{ijkl}}
\end{eqnarray}
where the Coulomb interaction matrix element, $J_{ijkl}$ , can be (i) the matrix element, $J_{ijij}$, for interaction between charges at $A$ and $B$, (ii) the matrix element, $J_{ijik} (J_{ijkj})$, for interaction between a charge at $A(B)$ with a transition density at $B(A)$, and (iii) the matrix element, $J_{ijkl}$, for interaction between transition densities at $A$ and $B$. The coupling, (iii) above, between the transition densities at $A$ and $B$ is dominant in the F\"orster limit and it mediates the motion of Frenkel excitons between subsystems. When the separation of subsystems $A$ and $B$ is large relative to the spatial extent of either subsystem the transition density coupling can be calculated accurately in the 
dipole-dipole limit \cite{ykhan}. The Exchange interaction matrix element, $K_{ijkl}$, accounts for the interactions between densities, $\rho_{ij} = a_ib_j$, at $A$ and $B$. The overlap between $a_i$ and $b_j$ decays exponentially with distance from the boundary of each subsystem. Therefore, the Exchange interaction can dominate only if the subsystems are very close to one another or if the singlet-singlet Coulomb interaction is symmetry forbidden. This constitutes the Dexter limit of energy transfer.

Thus
\begin{eqnarray}
    E_A 	&=&	{\rm Tr}\{\rho \mathcal H_A\}
	=	\sum_{ijkl} c_{ij}^{*}c_{kl}\langle \psi_{ij} \vert \Gamma_A \mathcal H \vert \psi_{kl} \rangle \nonumber \\
	&=&	\tfrac{1}{2} \sum_{ijkl} c_{ij}^{*}c_{kl} (E_{A_i}\delta_{ij,kl} + J_{ij,kl} - K_{ij,kl}) \nonumber \\
	&=& \sum_{ij} |c_{ij}|^2 E_{A_i} + \tfrac{1}{2} \sum_{ijkl} c_{ij}^{*}c_{kl} (J_{ij,kl} - K_{ij,kl})
\end{eqnarray}

Hence, the energy of subsystem $A$ is a sum of the energy
of the isolated molecule plus half of the coupling energy,
comprising the coulombic coupling $J$, and the exchange coupling $K$.
Since the interaction energy is split equally between the two subsystems
the total electronic coupling energy of the system
is twice the interaction energy assigned to subsystem $A$.
The electronic coupling matrix element is the sum of the Coulombic
coupling, J, and the Exchange coupling, K,
\begin{equation}
    V^{coul}	=	J + K~.
\end{equation}
These are the F\"orster and Dexter coupling results obtained by Parson
\cite{parson}.

\subsection{Multi-electron Case}
To extend the analysis of the previous section to
multi-electronic systems, let $\Pi({\bf M})$
be a single determinant M-electron wavefunction of
subsystem $A$.
\begin{equation}
  \vert \Pi(\mathbf{M}) \rangle   = \frac{1}{\sqrt{M!}}\hat A \vert a_{1}^{\Pi}(1)a_{2}^{\Pi}(2)\cdots a_{M}^{\Pi}(M) \rangle~,
\end{equation}
where the anti-symmetrization operator
\begin{equation}
	\hat A = \sum (-1)^{p} \hat P
\end{equation}
sums over all possible permutations $\hat P$ of the M electrons $1,2,\ldots,M$.
 The factor $(-1)^{p}$
represents the parity of the permutation $\hat P$. Similarly $\Gamma({\bf K})$
is the antisymmetric K-electron wavefunction of subsystem $B$.
\begin{equation}
  \vert \Gamma(\mathbf{K}) \rangle   = \frac{1}{\sqrt{K!}}\hat A \vert b^{\Gamma}_1(1)b^{\Gamma}_2(2)\cdots b^{\Gamma}_K(K) \rangle~.
\end{equation}
The antisymmetrized N-electron wavefunction, $\Psi({\bf N})$, of the composite system is then
\begin{equation}
  \vert \Psi(\mathbf{N}) \rangle = \left(\frac{K!(N-K)!}{N!}\right)^{1/2} {\hat F} \vert \Pi(\mathbf{M})\rangle \vert \Gamma(\mathbf{K})\rangle~,
\end{equation}
where $\hat F$ sums  over the permutations between the first M
and the remaining K electrons.  Then
\begin{eqnarray}
  \vert \Psi(\mathbf{N}) \rangle &= &\sqrt{\frac{K!(N-K)!}{N!}} {\hat F}
    \frac{1}{\sqrt{M!}}\hat A \vert a^{\Pi}_1(1)a^{\Pi}_2(2)\cdots a^{\Pi}_M(M) \rangle
    \frac{1}{\sqrt{K!}}\hat A \vert b^{\Gamma}_1(1)b^{\Gamma}_2(2)\cdots b^{\Gamma}_K(K) \rangle \nonumber \\
				   &=&  \sqrt{\frac{1}{N!}} {\hat F}
    \{\hat A \vert a^{\Pi}_1(1)a^{\Pi}_2(2)\cdots a^{\Pi}_M(M) \rangle\}
    \{\hat A \vert b^{\Gamma}_1(1)b^{\Gamma}_2(2)\cdots b^{\Gamma}_K(K) \rangle\}
\end{eqnarray}

In accord with our definition, the energy of subsystem $A$ is given by
\begin{eqnarray}
    E_A 	&=&	{\rm Tr}\{\rho \mathcal H_A\} \nonumber \\
		&=&	N \langle \Psi \vert h_A(1) \vert \Psi \rangle + \frac{N(N-1)}{2} \langle \Psi \vert g_{A,A}(1,2) \vert \Psi \rangle
		+ \frac{N(N-1)}{2} \sum_{\alpha \ne A} \langle \Psi \vert g_{A,\alpha}(1,2) \vert \Psi \rangle \nonumber \\
	&=& ~~~~ \sum_{i=1}^{M} \langle a_{i}(1) \vert h(1) \vert a_{i}(1) \rangle \nonumber \\
	&& +~~\sum_{\substack{i =1\\j>i}}^{M}\!^{'} \langle a_{i}(1)a_{j}(2) \vert g(1,2) \vert a_{i}(1)a_{j}(2) \rangle 
	 	- \langle a_{i}(1)a_{j}(2) \vert g(1,2) \vert a_{j}(1)a_{i}(2) \rangle \nonumber \\
	&& +~~ \sum_{i=1}^{M}\sum_{j=1}^{K}
		  \langle a_{i}(1)b_{j}(2) \vert g(1,2) \vert a_{i}(1)b_{j}(2) \rangle
		- \langle a_{i}(1)b_{j}(2) \vert g(1,2) \vert b_{j}(1)a_{i}(2) \rangle  \nonumber \\
	&=&	T_A + V_{A,A} + \frac{1}{2}V_{A,B}
\end{eqnarray}
In the limit where $A$ and $B$ are infinitely separated
 $V_{A,B} = 0$ so that $\lim_{R \to \infty} E_A = T_A + V_{A,A}$
we properly recover  the energy of the isolated subsystem $A$.

\section{Sample Computational Results}\label{sec:compute}
 We present numerical results for  coherent energy and charge transfer dynamics in several elementary systems:  (a) the
hydrogen molecule, $H_2$,  where each atom is considered as an
open subsystem, and (b) two hydrogen molecules interacting with one another.  Issues of nuclear antisymmetrization are neglected.
In the former case the interacting subsystems have open shell electronic configurations
while in the latter case the subsystems have closed shells.
Hartree-Fock (HF) molecular orbitals are obtained in both cases
using Gaussian 03 \cite{gaussian} at various bond lengths $R$, for which EET results are shown.  When the molecule is in a stationary vibrational state, EET results would be a weighted average over those shown below for various R values.  Despite the simplicity of the systems, the results are enlightening.

\begin{figure}
\begin{tabular}{ll}
a)&b)\\
\includegraphics[height=80mm,width=80mm]{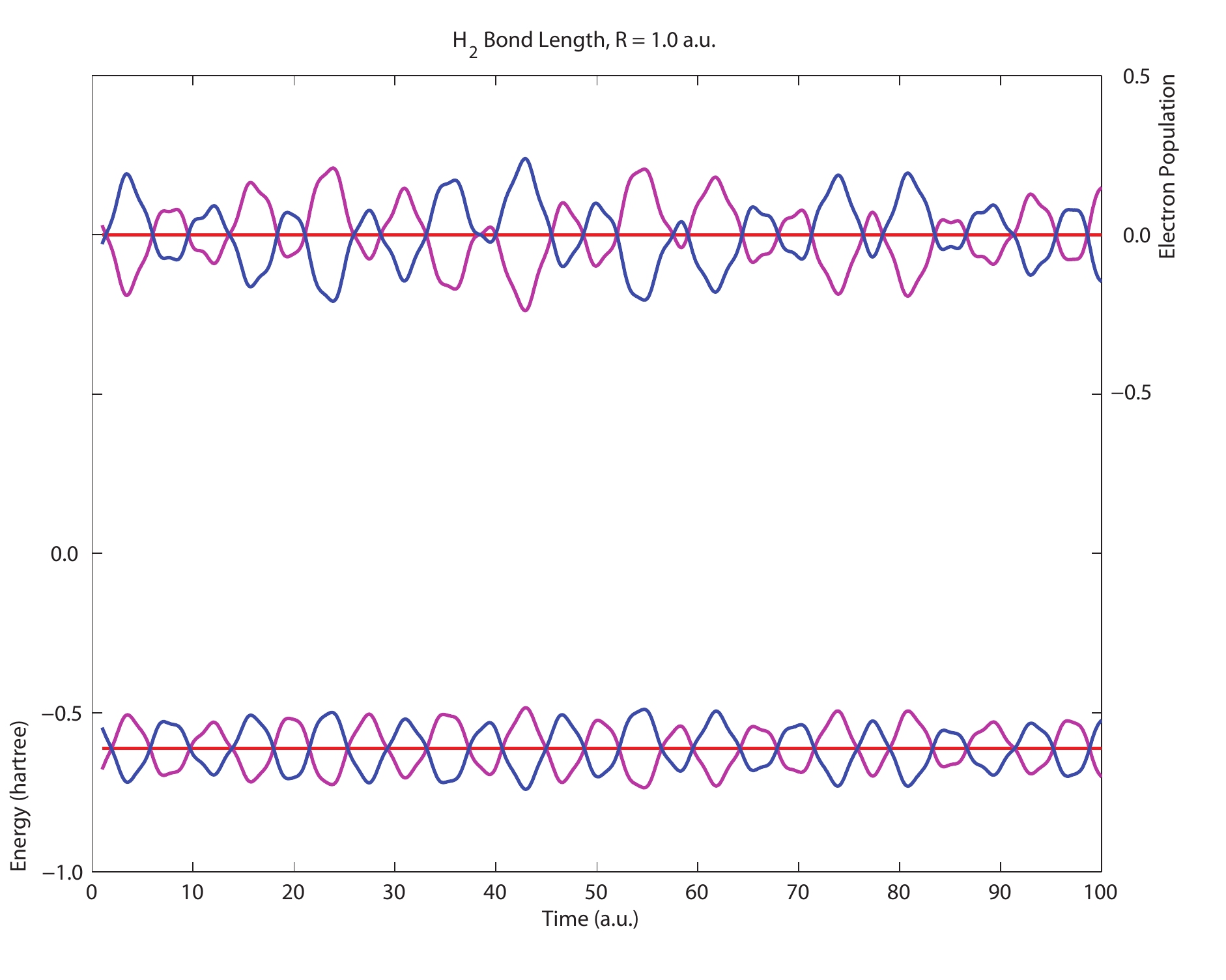}&
\includegraphics[height=80mm,width=80mm]{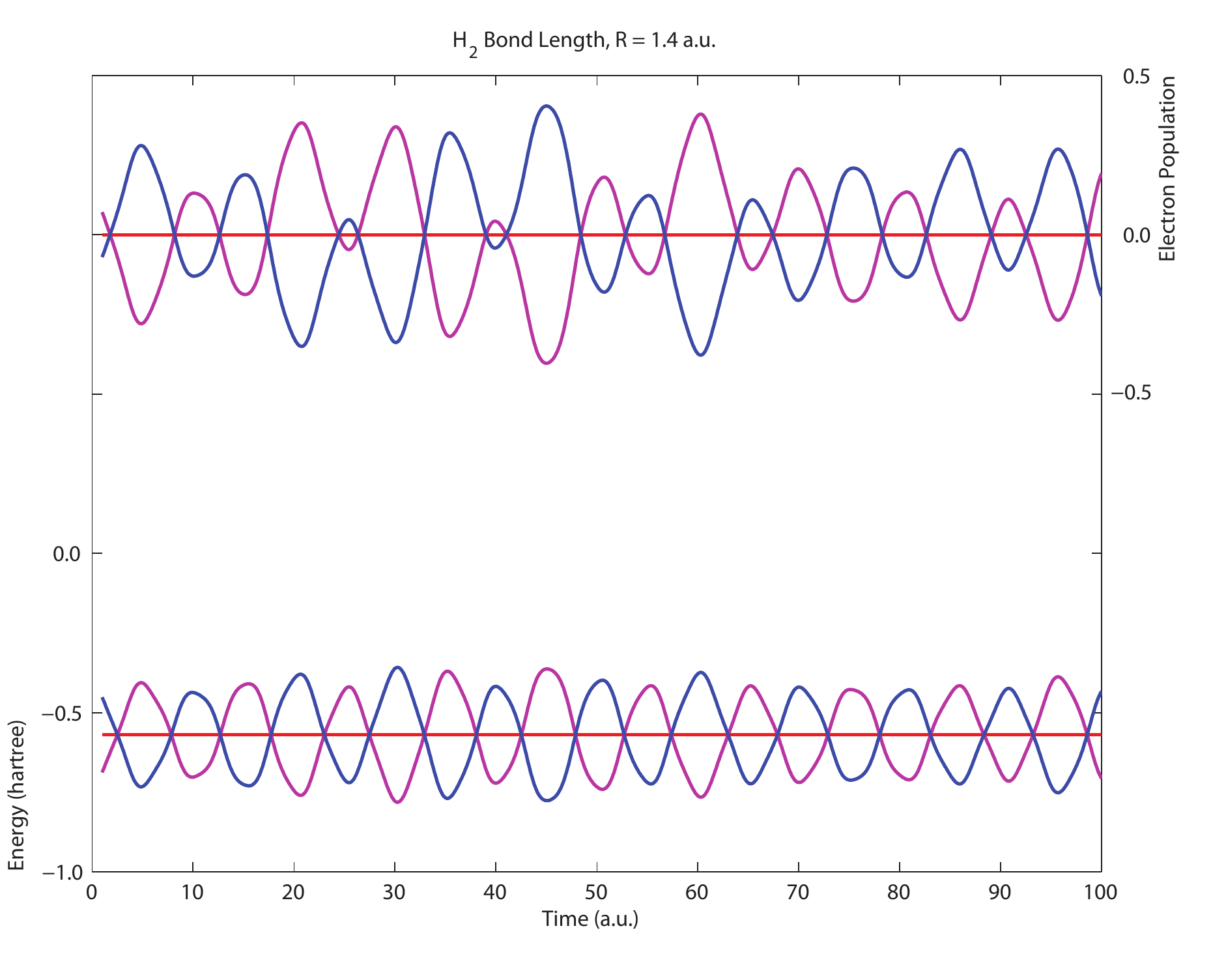}\\
c)&d)\\
\includegraphics[height=80mm,width=80mm]{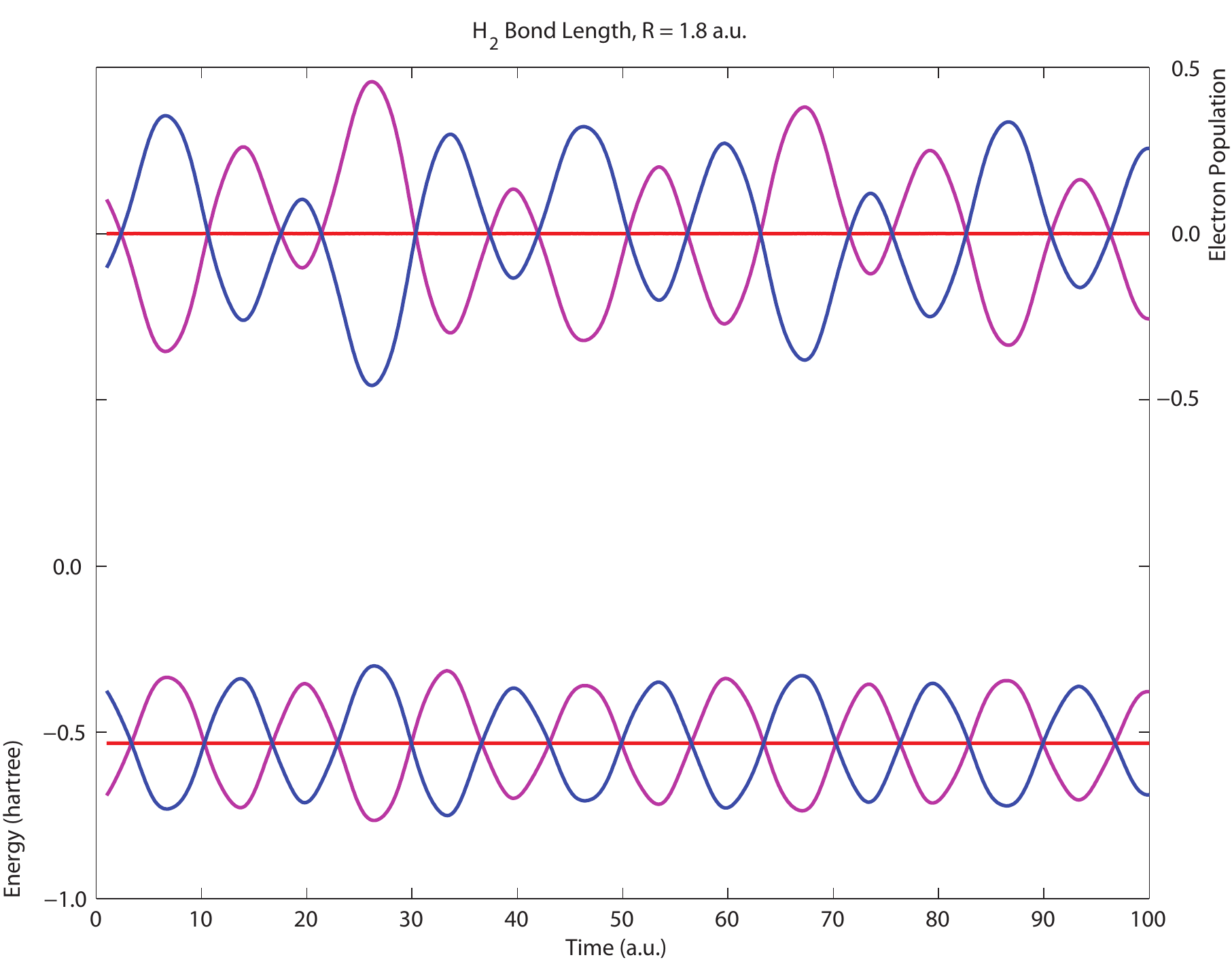}&
\includegraphics[height=80mm,width=80mm]{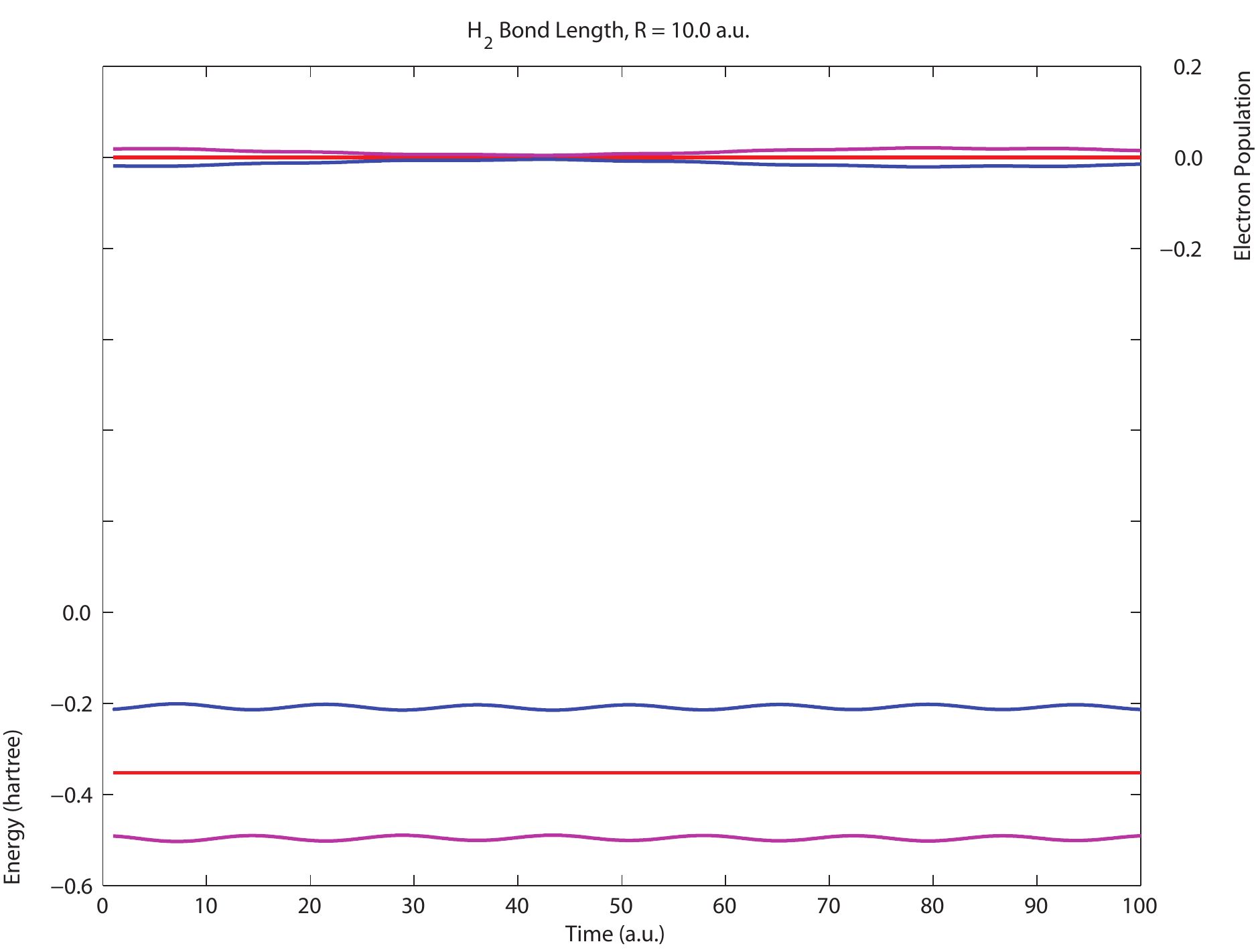}\\
\end{tabular}
\caption{$H_2$ molecule prepared in electronic state $|1s_A2p_{zB}\rangle$.
Coherent energy and charge transfer dynamics as function of time for bond lengths
(a) $R = 1.0 a.u.$, (b) $R = 1.4 a.u.$, (c) $R = 1.8 a.u.$, and (d) $R = 10.0 a.u.$}
\label{fig:spzResults}
\end{figure}

\subsection{Interacting Hydrogen Atoms}
 Figure~\ref{fig:spzResults} presents the coherent time evolution of energy and
population between two interacting hydrogen atoms initially in the state $|1s_A2p_{Z,B}\rangle$ at various fixed internuclear separations. In each panel
of the figure the curves
near the top show the time evolution of the population on each hydrogen atom, together
with their time average. The curves near the bottom of each panel show the
time evolution of the energy of each atom, together with their time average.

The figures confirm several intuitive expectations.
At large separation (Fig.~\ref{fig:spzResults}d), when the atoms are nearly
isolated,  there is negligible exchange of both energy and charge.
One hydrogen atom has the electronic ground state energy of $1s_A$,  while the other
hydrogen atom has the electronic excited state energy of $2p_B$. As the bond length
shortens, with an accompanied increase in the electronic coupling strength, we
observe the expected increase in the rate of energy transfer
(Fig.~\ref{fig:spzResults}(a-c)).

\begin{figure}
\begin{tabular}{ll}
a)&b)\\
\includegraphics[height=80mm,width=80mm]{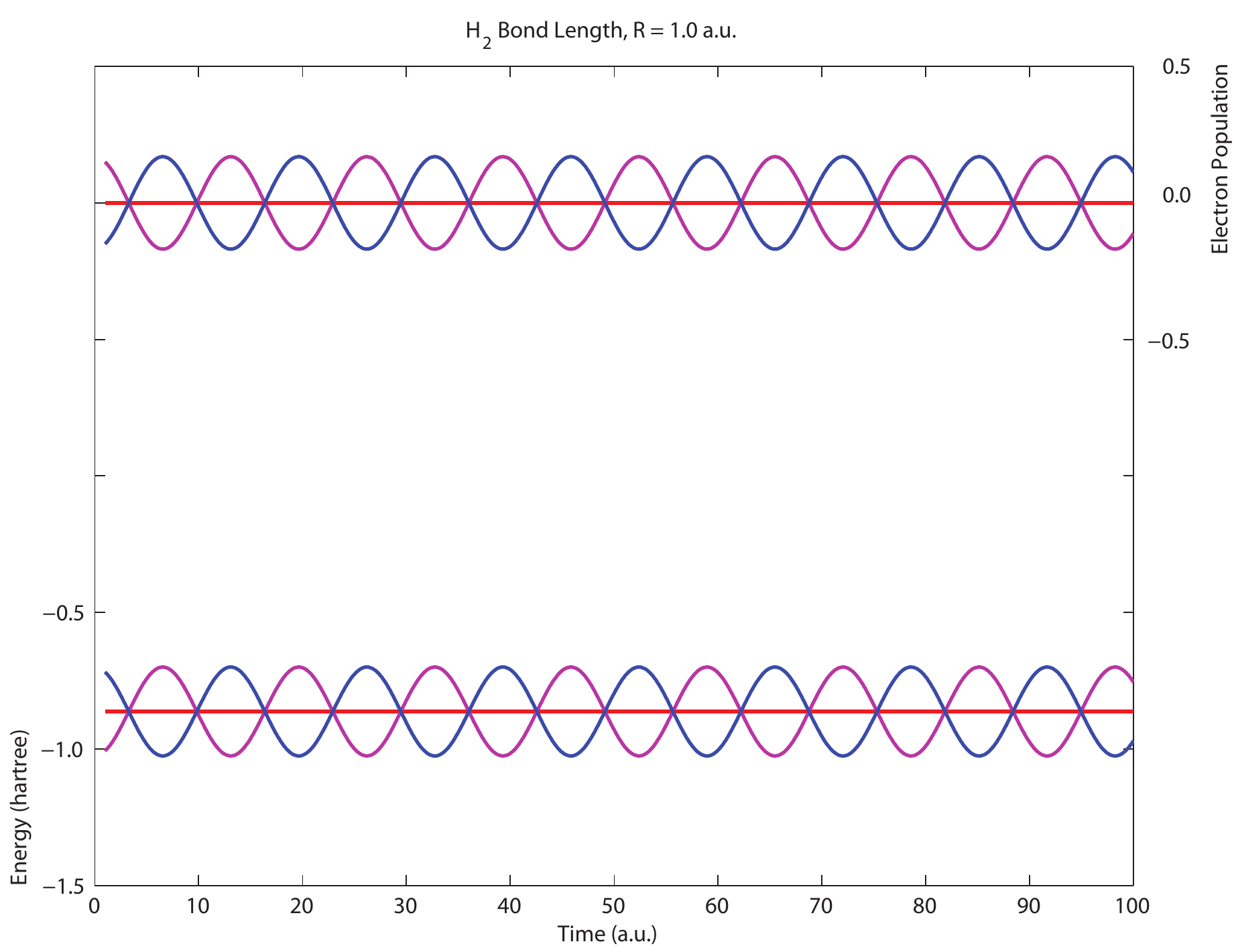}&
\includegraphics[height=80mm,width=80mm]{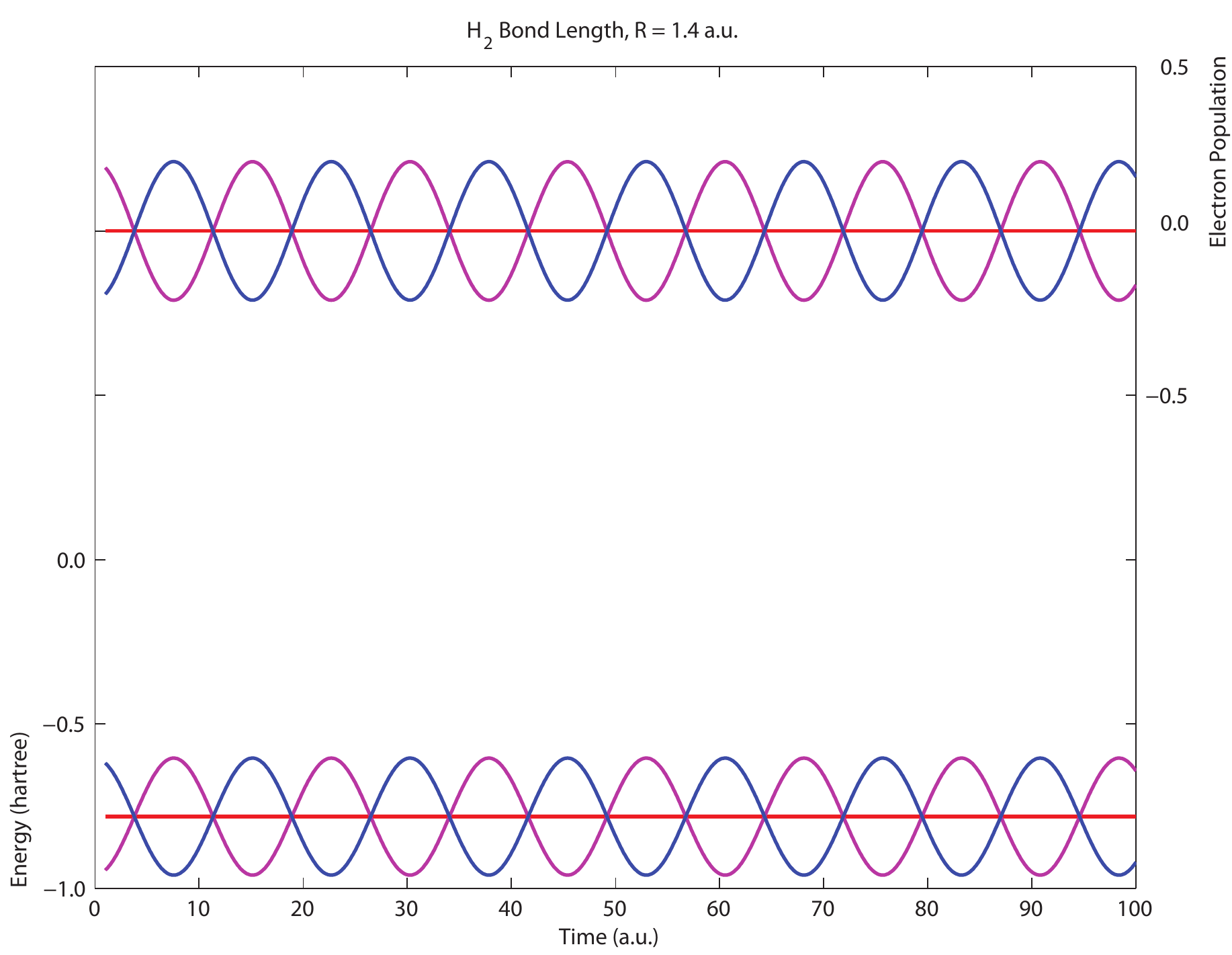}\\
c)&d)\\
\includegraphics[height=80mm,width=80mm]{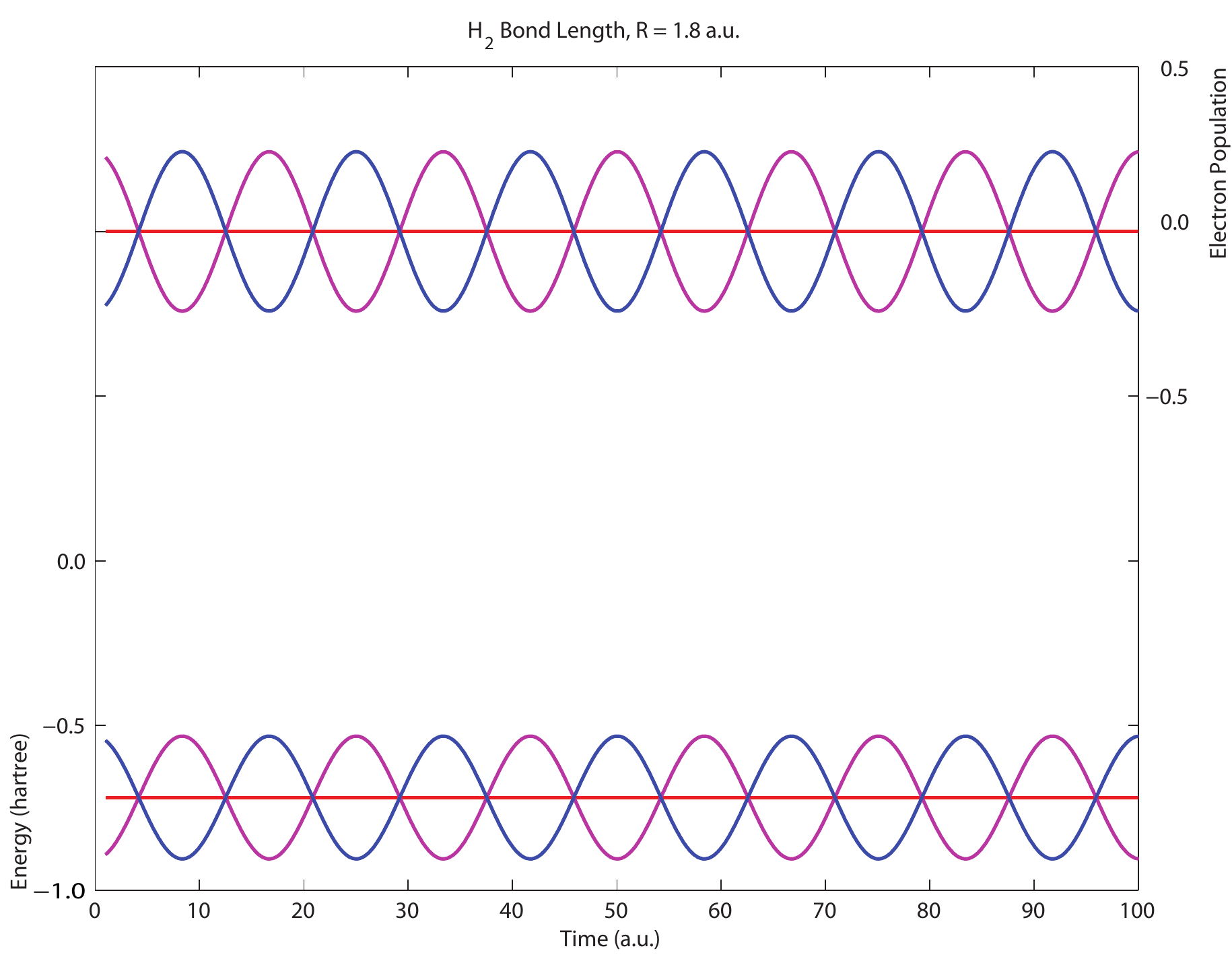}&
\includegraphics[height=80mm,width=80mm]{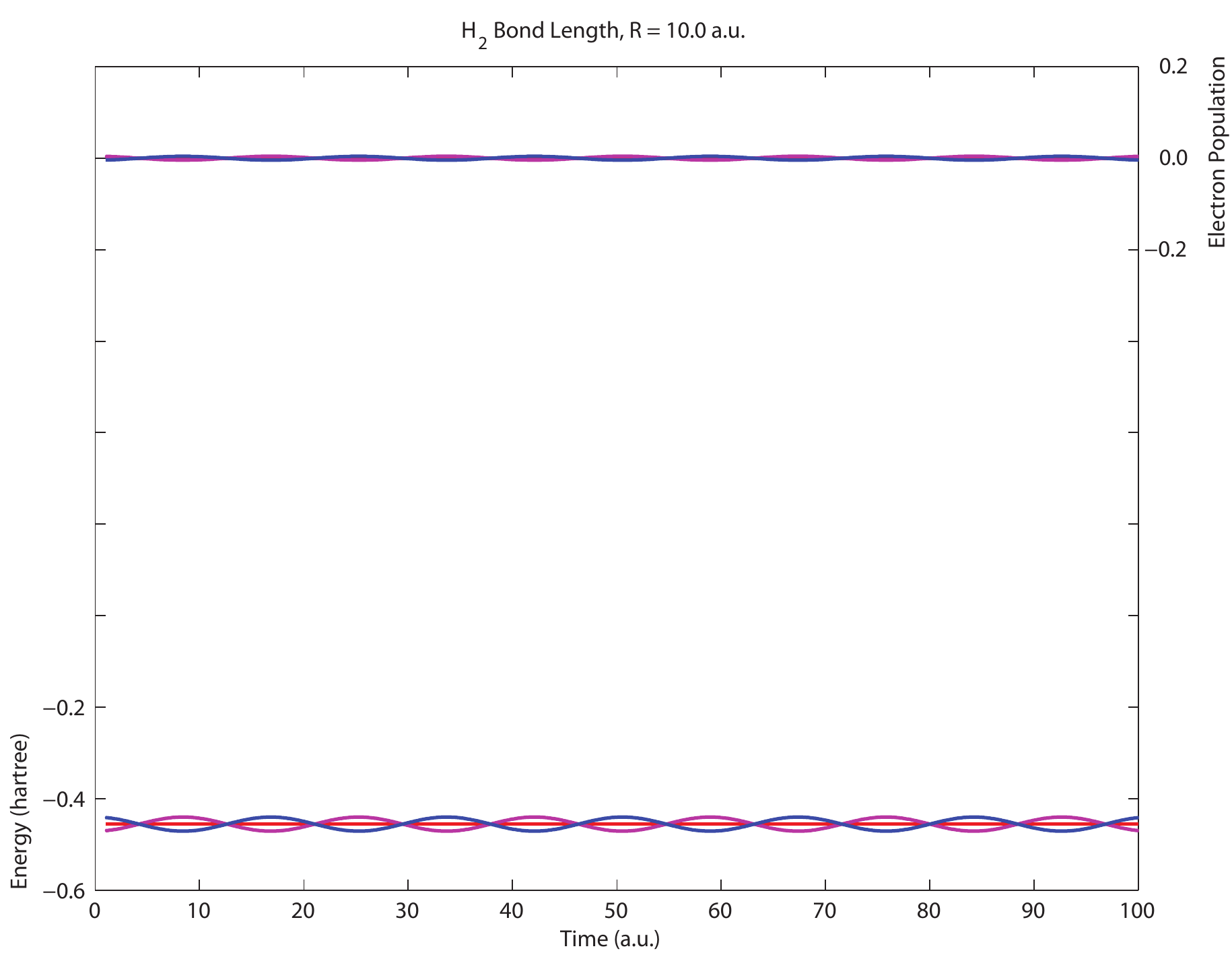}\\
\end{tabular}
\caption{$H_2$ molecule prepared in electronic superposition state
$|\Psi\rangle = 1/\sqrt{2} (|\phi_{ground}\rangle+|\phi_{1^{st} excited}\rangle) \approx
1/\sqrt{2}(|\sigma_{1s}\sigma_{1s}\rangle + |\sigma_{1s}\sigma_{1s}^{*}\rangle)$.
Coherent energy and charge transfer dynamics as function of time for bond lengths
(a) $R = 1.0 a.u.$, (b) $R = 1.4 a.u.$, (c) $R = 1.8 a.u.$, and (d) $R = 10.0 a.u.$}
\label{fig:e1e2Results}
\end{figure}

The figure also
reveals two interesting features. First,  energy transfer is here seen to be perfectly
correlated with charge transfer for all $R$. Also,  the amount of energy and charge exchanged coherently between the hydrogen
atoms  maximizes at some intermediate separation, consistent with the fact that 
 no energy or charge will be exchanged in the two limits, i.e., in the limit of the
unified atom and that of infinitely separated atoms.

To further illustrate these patterns,  consider next a hydrogen molecule prepared
in a superposition of its ground and first excited electronic state
(Fig.~\ref{fig:e1e2Results}). The dominant contribution to the ground electronic state
is from the $\sigma_{1s}^{2}$ configuration.
The first excited electronic state has a primarily $\sigma_{1s}\sigma_{1s}^{*}$
character, where $\sigma_{1s} \sim (1s_A + 1s_B)$ is the sigma bonding molecular orbital
while $\sigma^{*}_{1s} \sim (1s_A - 1s_B)$ is the sigma anti-bonding molecular orbital.
A superposition of these states gives rise to coherent
dynamics between the $1s$ orbitals on the two hydrogen atoms.  Here too, the rates of charge and energy transfer as the two
hydrogen atoms approach each other increase, and there is  an optimal $R$ at which  energy and charge transferred are maximized. At infinite separation limit the two
hydrogen atoms have equal energy, corresponding to that of  isolated $|1s\rangle$ hydrogen atoms.

The dynamics in this case, which    involves the superposition of the two
lowest singlet electronic states look much simpler than that in Fig. \ref{fig:spzResults}. By contrast, the state in  Fig. \ref{fig:spzResults}   projects
onto several CI eigenstates of the H$_2$ molecule and the dynamics therefore involves
multiple timescales corresponding to many eigenenergy differences.

\subsubsection*{Decoherence}
As a further example we would like to consider the effect of decoherence, e.g., decoherence
induced by collisions of gaseous $H_2$ molecules, on the coherent energy and charge
transfer dynamics. We assume that the collisions as elastic, resulting in a dephasing of
otherwise coherent dynamics. Moreover, we write the complete wavefunction of our
system as a product of the electronic and nuclear wavefunctions in the
Born-Oppenheimer approximation. For the superposition state studied above
(Fig.~\ref{fig:e1e2Results}), we may write the wavefunction as:
\begin{eqnarray}
	|\Psi(r,R)\rangle &=& \frac{1}{\sqrt{2}}|\Psi_g(r,R)\rangle + \frac{1}{\sqrt{2}}|\Psi_e(r,R)\rangle  \nonumber \\
	& =& \frac{1}{\sqrt{2}}|\psi_g(r;R)\rangle|\chi_g(R)\rangle + \frac{1}{\sqrt{2}}|\psi_e(r;R)\rangle|\chi_e(R)\rangle
\end{eqnarray}
where $|\chi_g(R)\rangle $ and $|\chi_e(R)\rangle$ are taken to be the ground 
vibrational wavefunctions on the respective ground and excited electronic
potential energy surfaces (Fig~\ref{fig:potential}). The electronic energy on hydrogen
 atom $A$ is then:
\begin{figure}
\includegraphics[height=80mm,width=80mm]{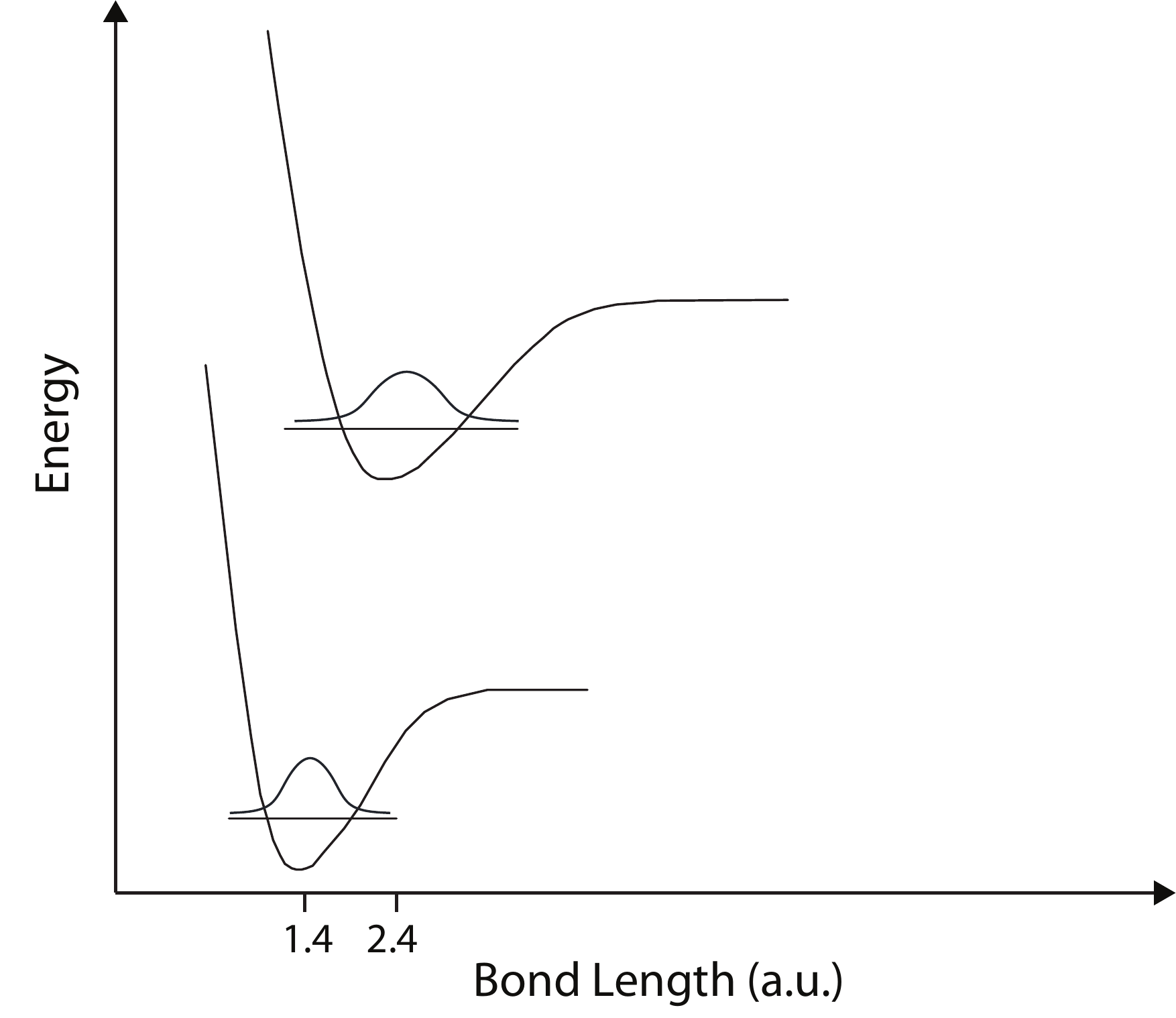}
\caption{A sketch of the ground and excited electronic potential energy surfaces
of the $H_2$ molecule}
\label{fig:potential}
\end{figure}
\begin{eqnarray}
	\langle E_A \rangle &=& \langle \Psi(r,R) |\mathcal{H}_A| \Psi(r,R)\rangle  \nonumber \\
		&=& 	\frac{1}{2} \sum_{i=g,e}\int dR \chi_i^{*}(R) \chi_i(R) \left( \int dr \psi_i^{*}(r;R) \mathcal{H}_A \psi_i(r;R)\right) \nonumber \\
	&&	+~~ \Re  \int dR \chi_g^{*}(R) \chi_e(R) \left( \int dr \psi_g^{*}(r;R) \mathcal{H}_A \psi_e(r;R) \right)  
	\end{eqnarray}

In order to make the expression more concise we define the following quantities
\begin{eqnarray}
		\omega_i	&=& E_i + (n_i+ \frac{1}{2})h\nu_i \nonumber \\
	\omega_{eg} &=& (E_e - E_g) +  (n_e + \frac{1}{2})h\nu_e  - (n_g + \frac{1}{2})h\nu_g \nonumber \\
	E_{Ai} &=& \frac{1}{2} \int dR \chi_i^{*}(R) \chi_i(R) \left( \int dr \psi_i^{*}(r;R) \mathcal{H}_A \psi_i(r;R)\right) \nonumber \\
		\Delta_A &=& \Re  \int dR \chi_g^{*}(R) \chi_e(R) \left( \int dr \psi_g^{*}(r;R) \mathcal{H}_A \psi_e(r;R) \right) \end{eqnarray}
where $i = g,e$.  Moreover, the time dependence of the wavefunction is
easily included; only the cross term $\Delta_A$  will be time dependent. Thus,  the time dependent energy on Hydrogen atom $A$ in the concise form
$$
\begin{array}{lll}
	\langle E_A(t) \rangle &=& 1/2 \left( E_{Ag} + E_{Ae} \right) + \Delta_{A}\cos(\omega_{eg}t)
\end{array}
$$
The net result is a time-dependent site electronic energy, reflecting the time-dependent superposition state.

\begin{figure}
\begin{tabular}{ll}
a)&b)\\
\includegraphics[height=80mm,width=80mm]{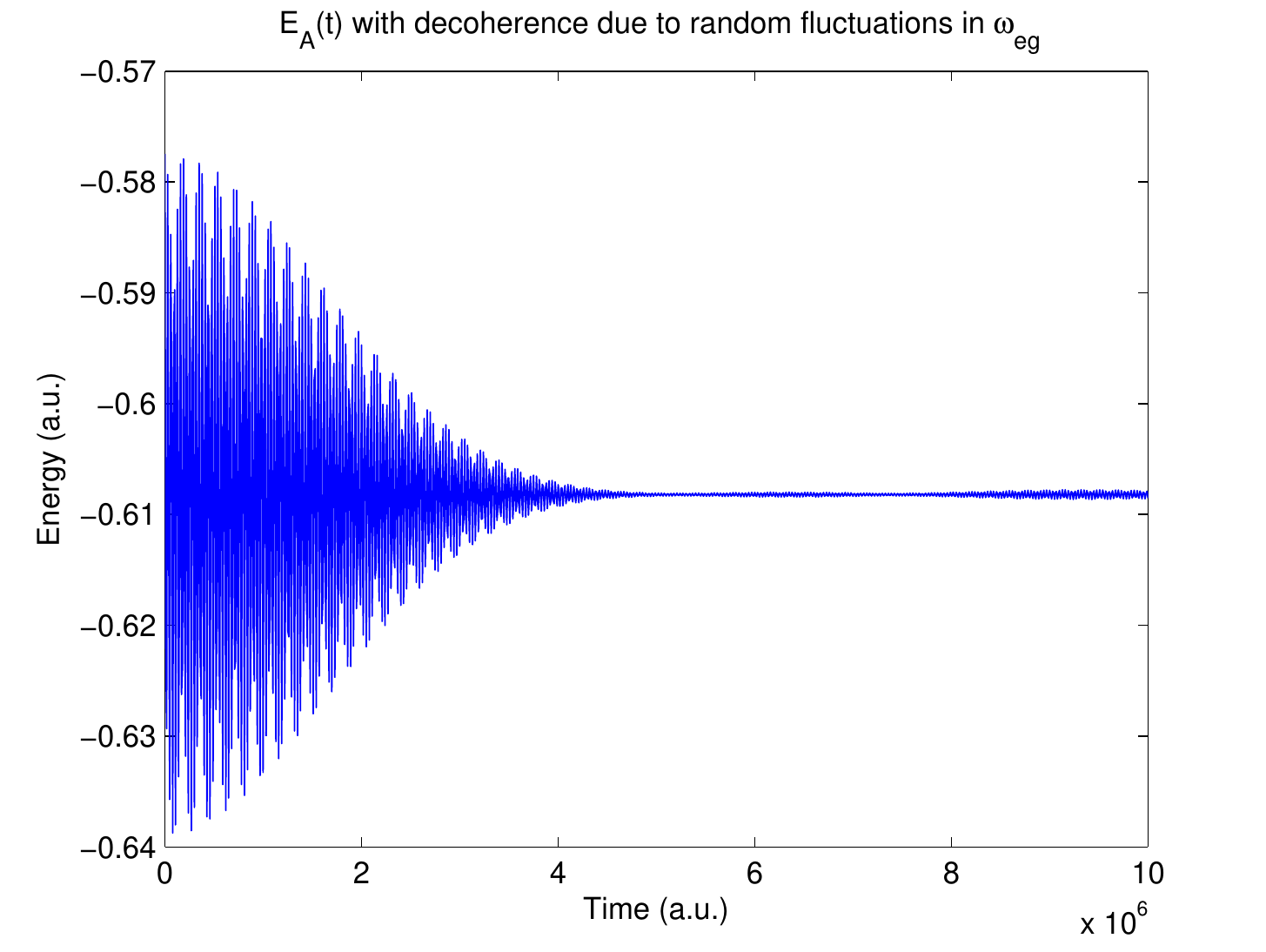} &
\includegraphics[height=80mm,width=80mm]{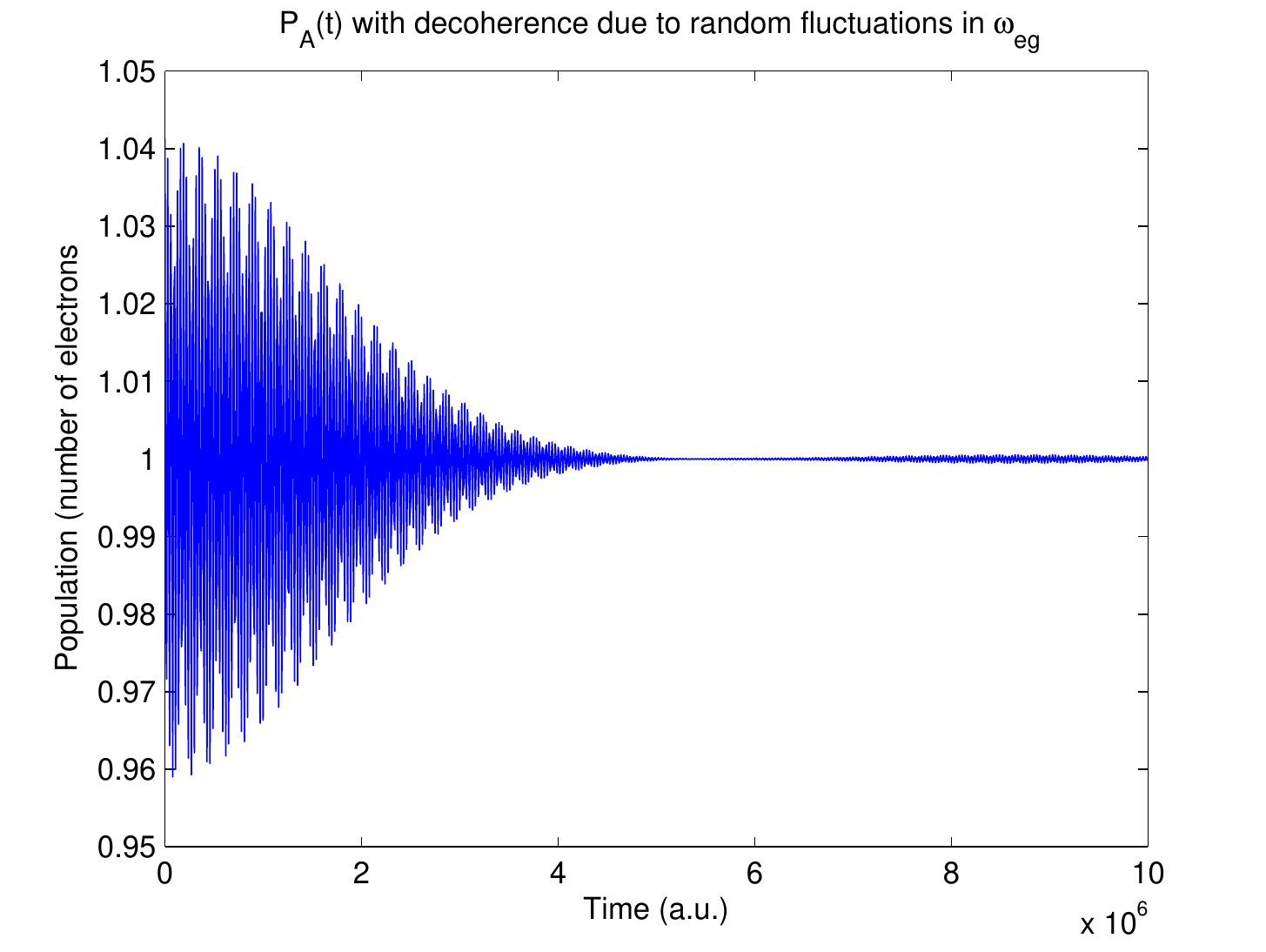} \\
\end{tabular}
\caption{Decoherence induced dynamics in (a) Energy, and (b) Charge of the
the $H_2$ molecule prepared in a superposition of the two lowest electronic states}
\label{fig:decohere}
\end{figure}

In addition, in the presence of a collisional environment the vibronic energy levels of the
system are expected to fluctuate about their mean positions leading to different
Hydrogen atoms in an ensemble of molecules acquiring an arbitrary phase.
The expectation value of the time-dependent energy on hydrogen atom $A$ can then be expressed as 
$$
\begin{array}{lll}
	\overline{\langle E_A(t) \rangle} &=&
	\frac{1}{2} \left( E_{Ag} + E_{Ae} \right) + \Delta_{A} \int d\omega \cos(\omega t)f(\omega)
\end{array}
$$
Here $f(\omega)$ describes the distribution in energies due to environmental collisions.
We assume a normal distribution of phases with mean $\omega_{eg}$.  In our simulations the
standard deviation is taken to be  $10^{-6}$ Hartree (corresponding to a
collision time on the order of 25 ps). The results of the simulation are shown in
Figure~\ref{fig:decohere}. We observe that for the given parameters the electronic energy transfer dynamics fades out on a timescale of $4\times10^{6}$ a.u. (100 ps).

\subsection{Interacting Hydrogen Molecules}
 As a second example,  consider the energy and population transfer dynamics between two interacting
Hydrogen molecules at various separations.
 As an example, we fix the bond distance within each hydrogen molecule at the equilibrium bond length of $1.4 a.u.$ and vary the distance
between the centers of the two H$_2$ molecules. For each nuclear geometry the Hartree-Fock molecular orbitals
of the system are determined using the 6-31G basis set, and the singlet excited states of the system
are determined at the CI-Singles (CIS) level.

\begin{figure}
\begin{tabular}{ll}
a)&b)\\
\includegraphics[height=80mm,width=80mm]{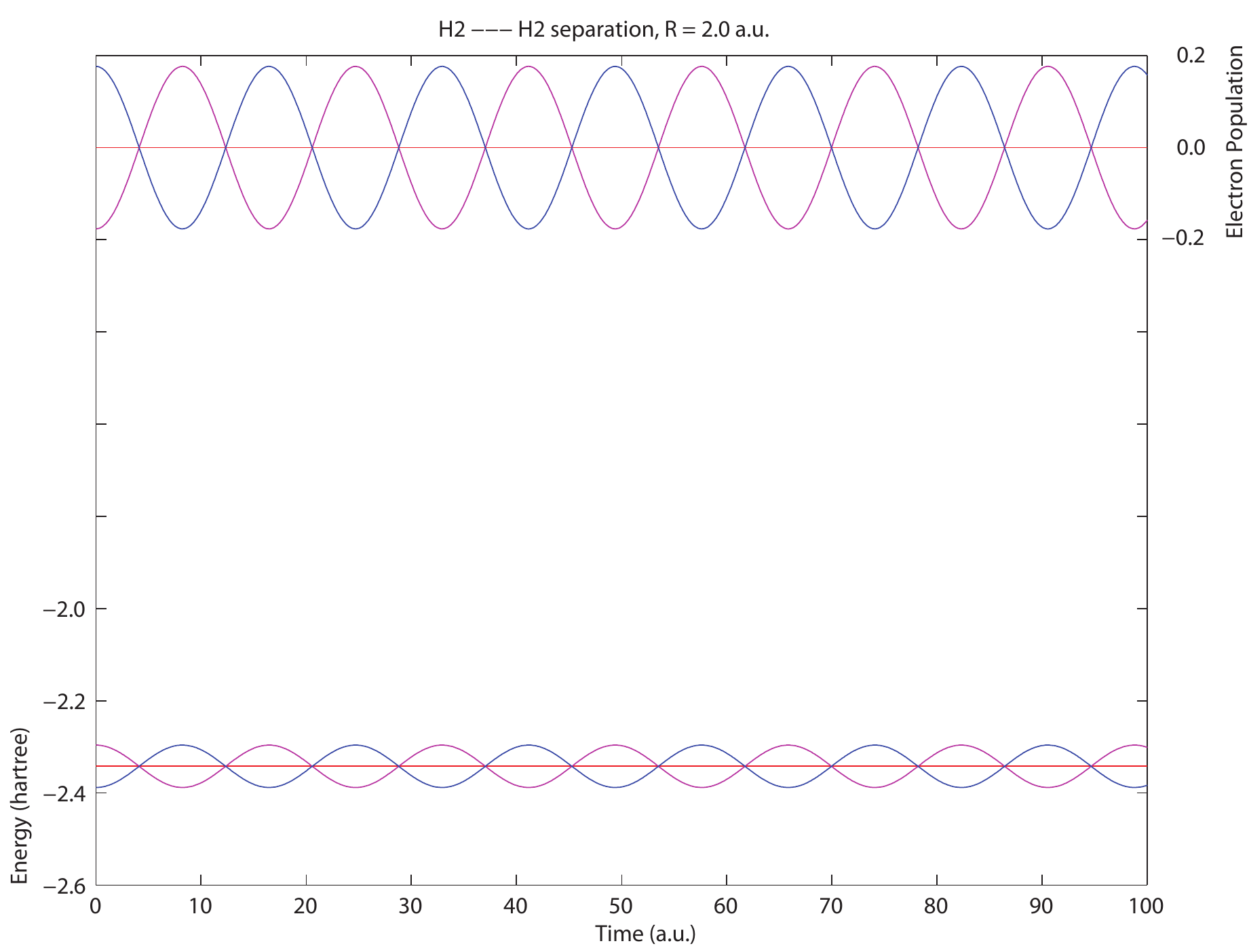}&
\includegraphics[height=80mm,width=80mm]{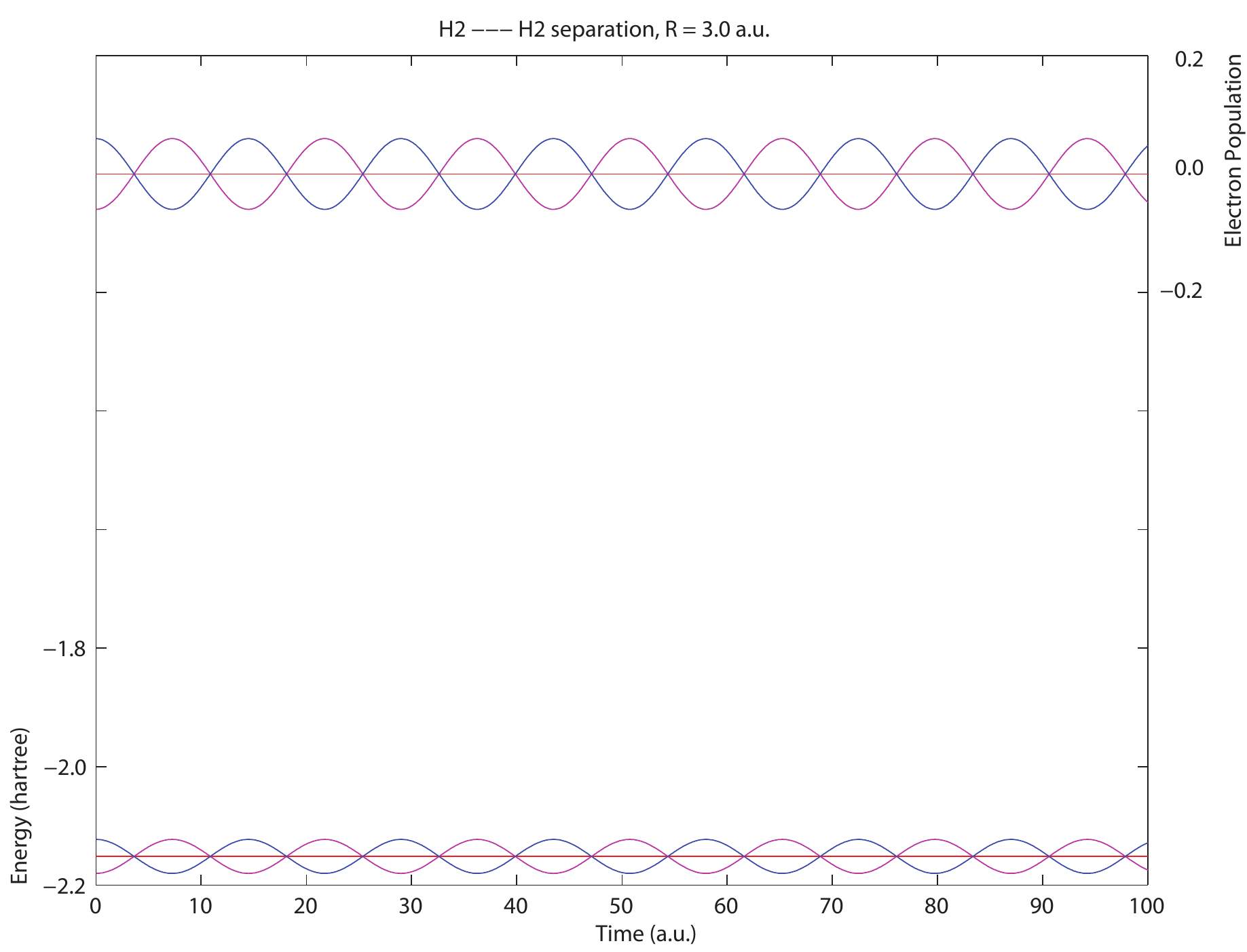}\\
c)&d)\\
\includegraphics[height=80mm,width=80mm]{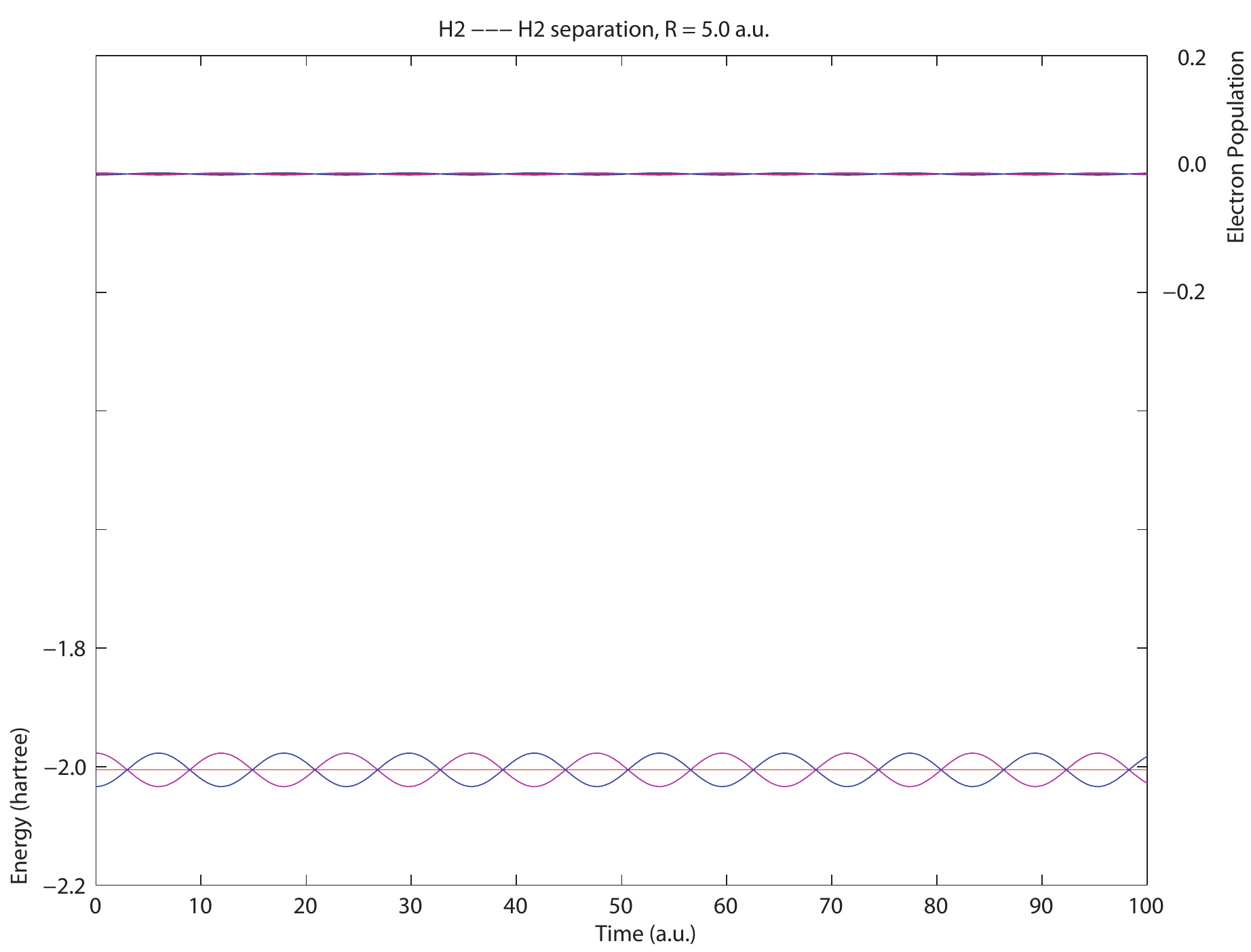}&
\includegraphics[height=80mm,width=80mm]{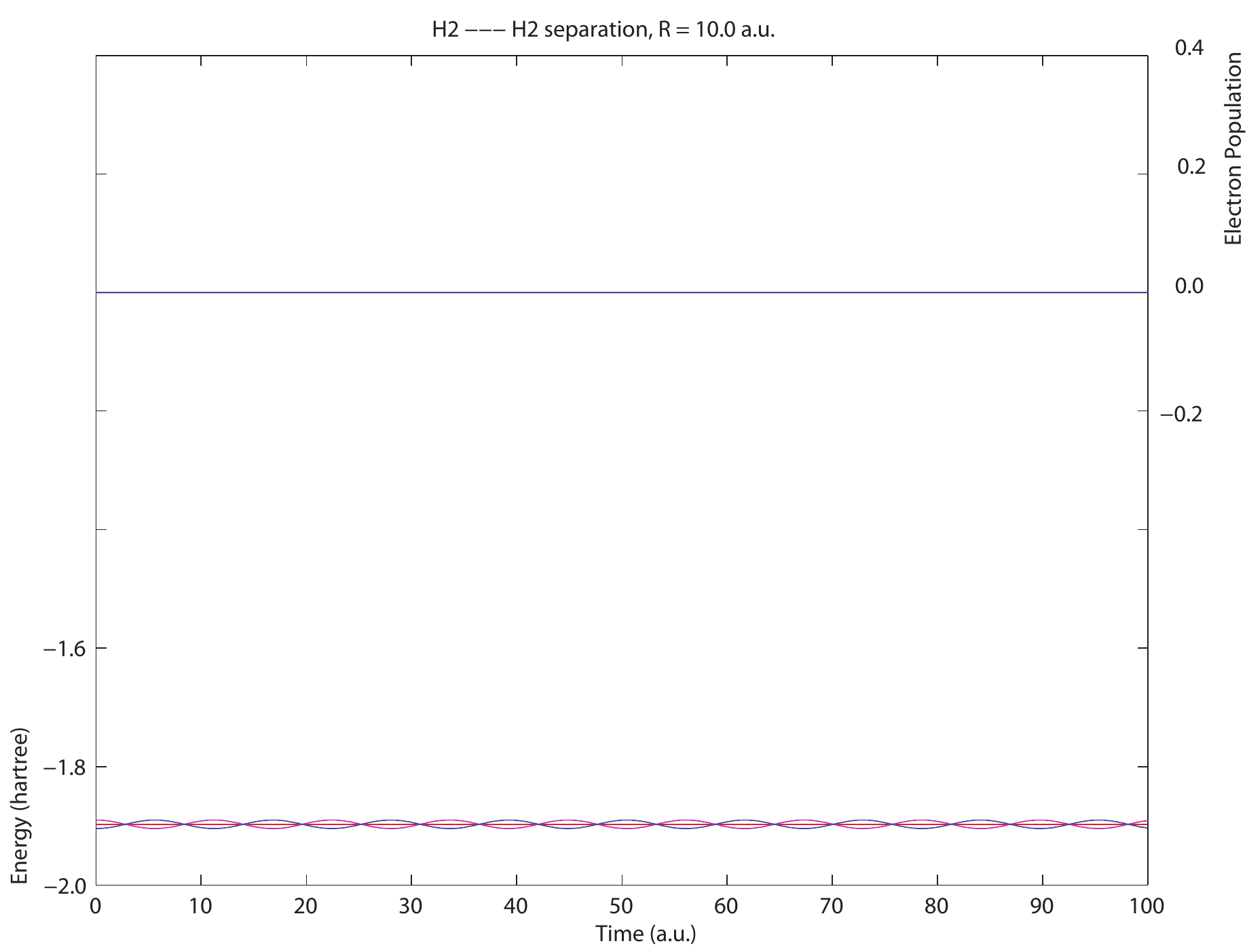}\\
\end{tabular}
\caption{Interacting $H_2$ molecules prepared in an electronic superposition state
of the ground and first excited CIS state.
$|\Psi\rangle = 1/\sqrt{2} (|\phi_{ground}\rangle+|\phi_{1^{st} excited}\rangle)$.
Coherent energy and population transfer dynamics as function of time for bond lengths
(a) $R = 2.0 a.u.$, (b) $R = 3.0 a.u.$, (c) $R = 5.0 a.u.$, and (d) $R = 10.0 a.u.$}
\label{fig:intmolp2}
\end{figure}

\begin{figure}
\begin{tabular}{ll}
a)&b)\\
\includegraphics[height=80mm,width=80mm]{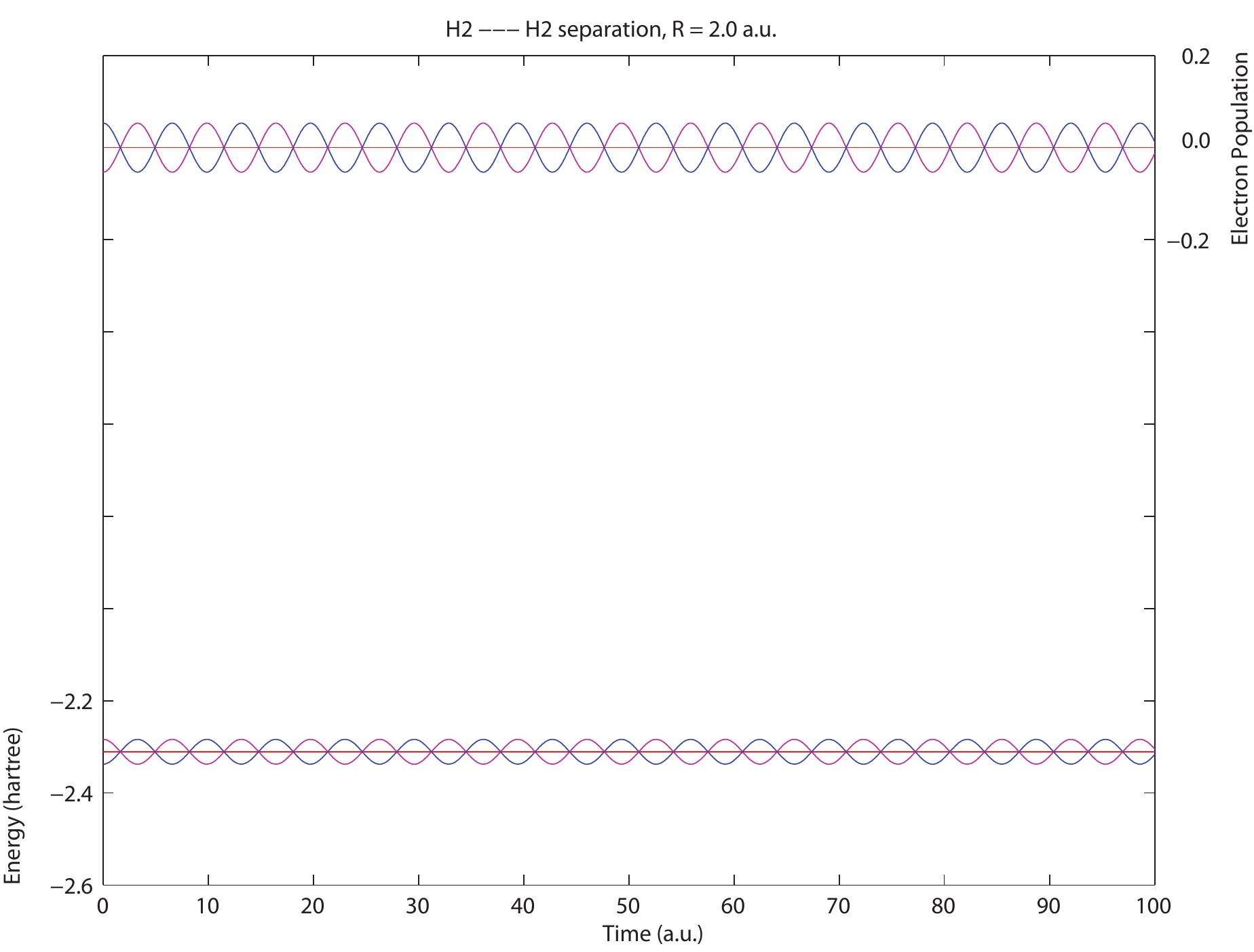}&
\includegraphics[height=80mm,width=80mm]{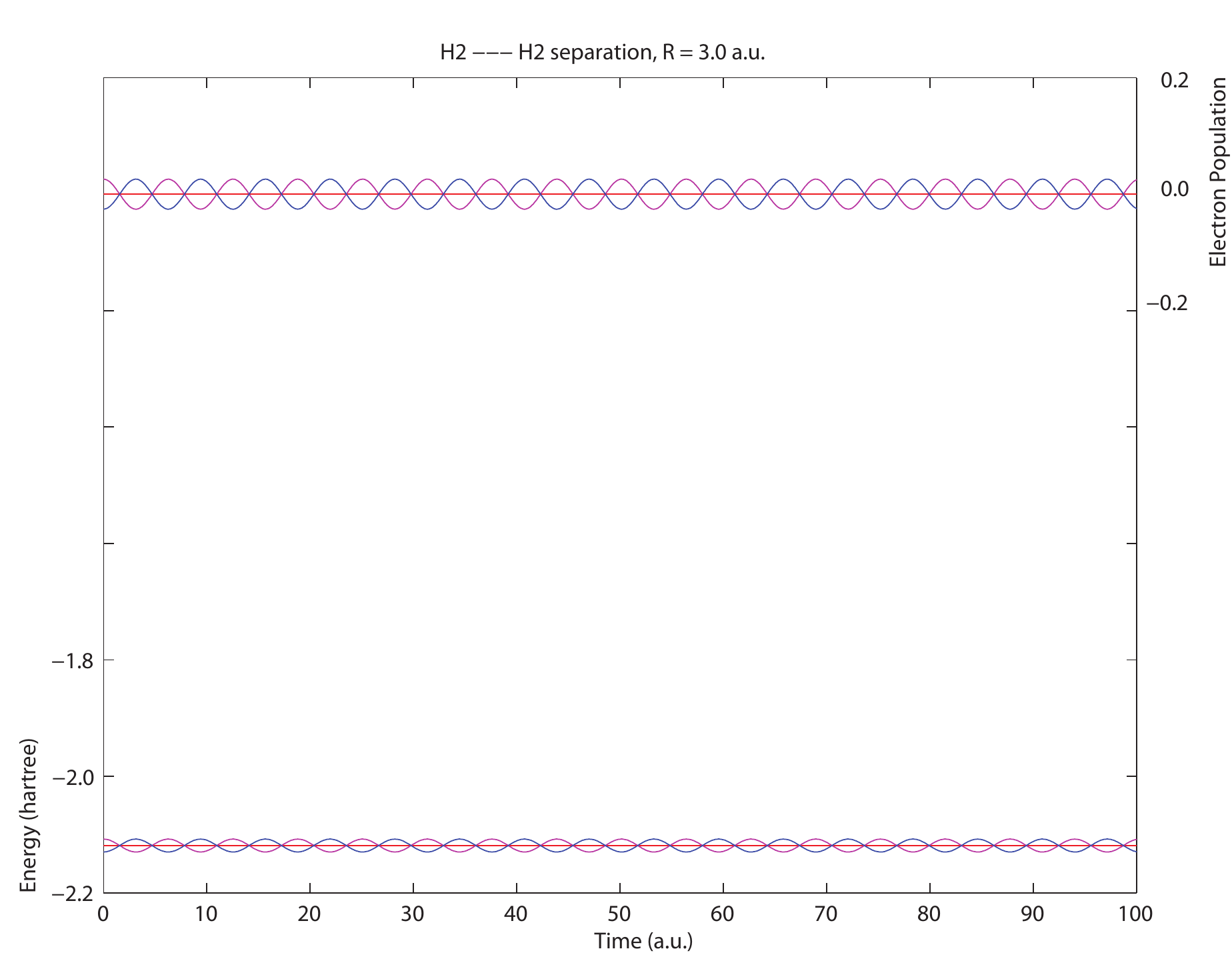}\\
c)&d)\\
\includegraphics[height=80mm,width=80mm]{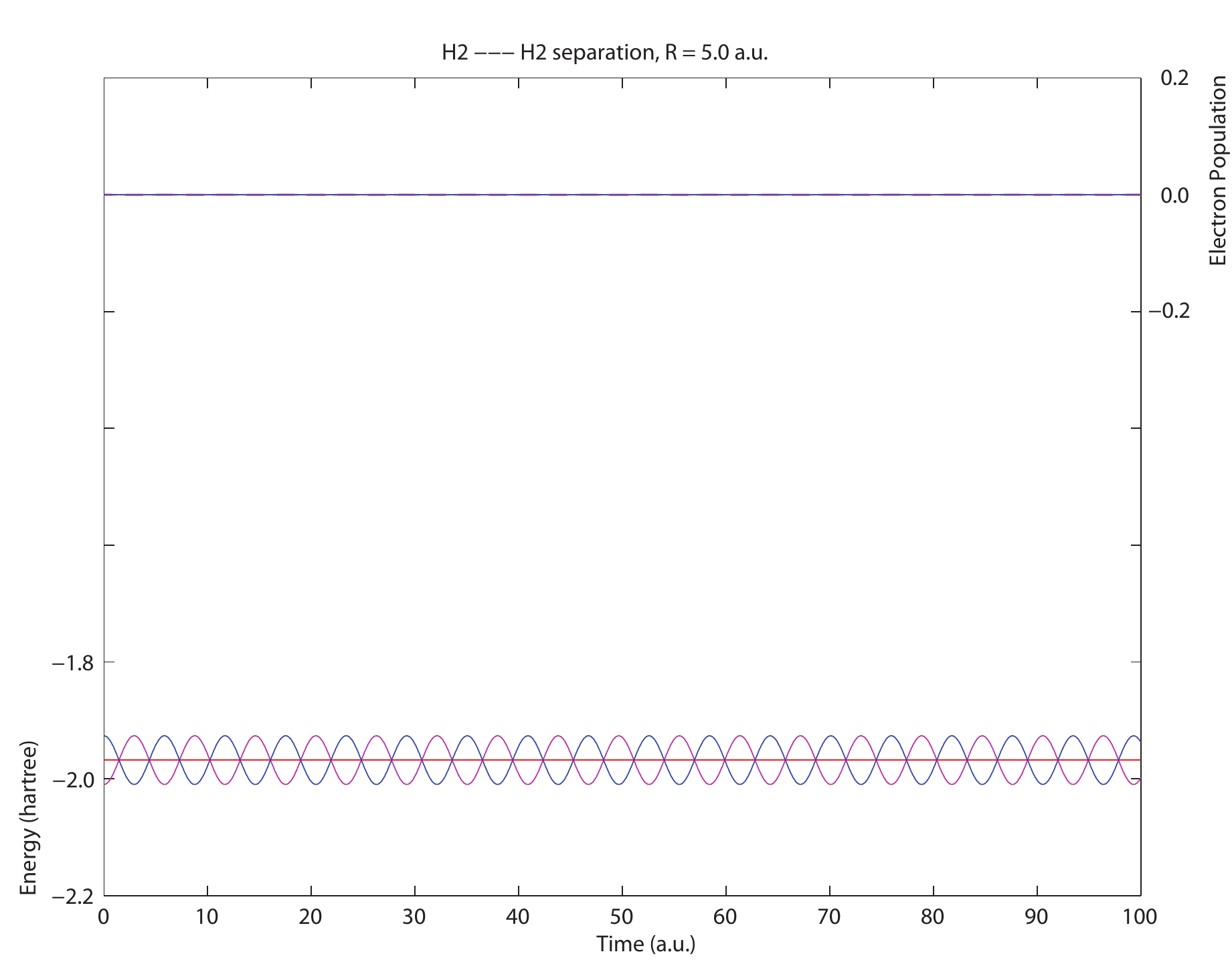}&
\includegraphics[height=80mm,width=80mm]{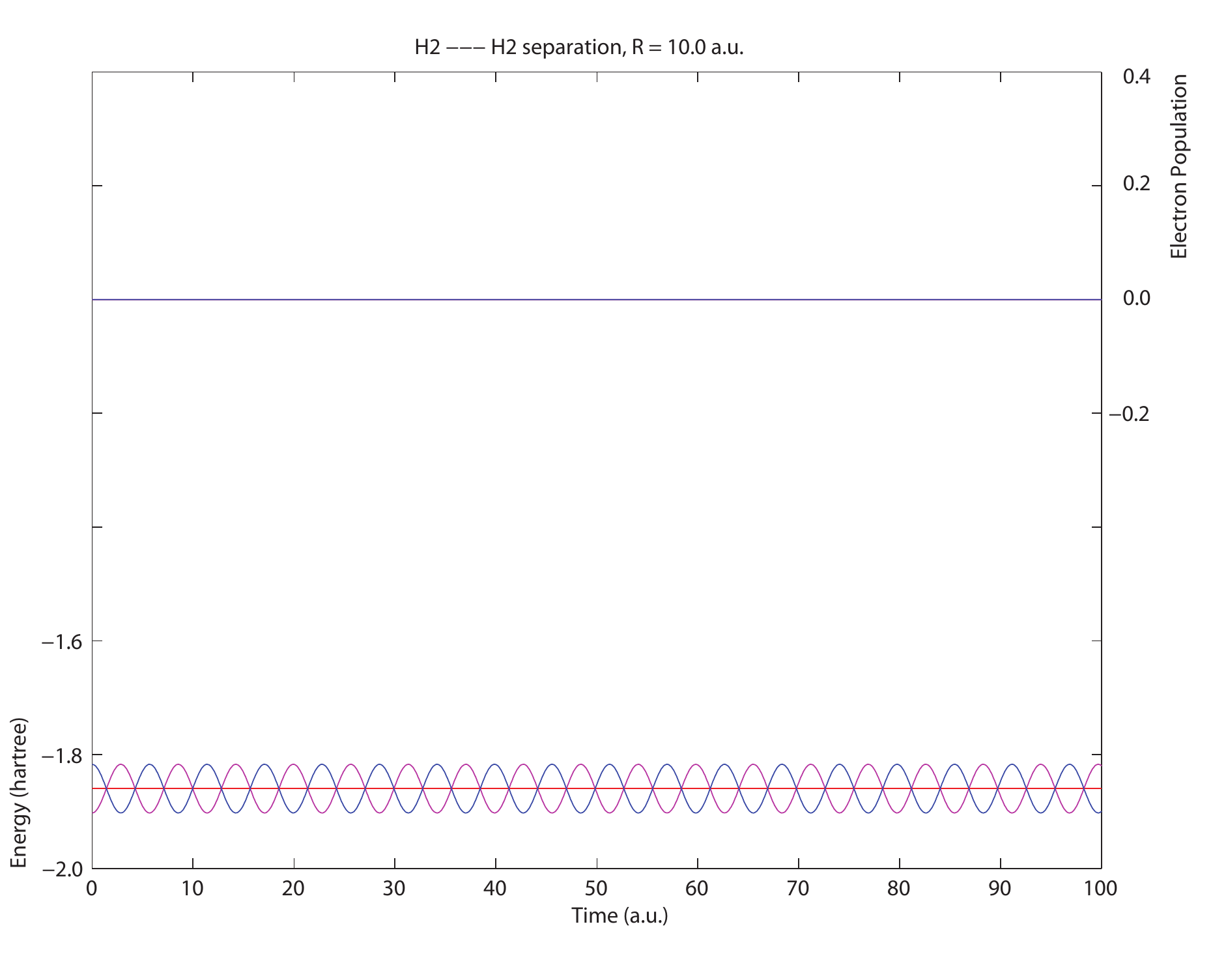}\\
\end{tabular}
\caption{Interacting $H_2$ molecules prepared in an electronic superposition state
of the ground and third excited CIS state.
$|\Psi\rangle = 1/\sqrt{2} (|\phi_{ground}\rangle+|\phi_{3^{rd} excited}\rangle)$.
Coherent energy and population transfer dynamics as function of time for bond lengths
(a) $R = 2.0 a.u.$, (b) $R = 3.0 a.u.$, (c) $R = 5.0 a.u.$, and (d) $R = 10.0 a.u.$}
\label{fig:intmolp4}
\end{figure}

Figure~\ref{fig:intmolp2} presents the coherent time evolution of energy and population
between the interacting molecules when the system is initiated in a superposition of its
ground and first excited singlet CIS electronic state. Similarly, Figure~\ref{fig:intmolp4}
shows the dynamics for a superposition of the ground and a higher lying electronic excited state.
The curves for the evolution of population and energy are respectively near the top and
bottom of each panel.

The dynamics of the interacting closed shell subsystems can be compared to the dynamics of the
interacting open shell system presented earlier. As before, the amplitude of population and
energy exchanged between the interacting subsystems decays with increasing separation. The
correlation between the population and energy dynamics within each subsystem, and the anticorrelation
between the dynamics of the two subsystems also persists. However, two important differences can be discerned.
First,  note that charge transfer is only significant at separations below 5.0 a.u., where as
significant energy transfer may persist beyond 10 a.u. This is in contrast to the two hydrogen atoms
where we found that charge transfer always accompanied energy transfer. Moreover, because we no
longer have the unified atom limit, the amplitude of energy and population transfer no longer shows
a clear trend at small separations.

\section{Summary}
\label{sec:conclusion}
We have introduced a well defined operator $\mathcal{H}_A$ 
that allows, given a time dependent or time independent wavefunction 
$|\Psi(t)\rangle$, the computation of electronic energy on a local site 
in a composite molecular system, as 
$E_A = \langle \Psi(t)|\mathcal H_A |\Psi(t)\rangle$. The definition
resolves numerous problems arising from electron interchange 
antisymmetrization that exist with naive approaches. Further, the
computation of $E_A$, 
being an integral over an operator that is a function of the electronic
degrees of freedom,  automatically includes decoherence effects due to 
other degrees of freedom in the molecule. The resultant operator has
been shown to give appropriate results in various limits and to provide
insight into electronic energy dynamics in small molecular systems.
Applications to larger systems are underway.

\textbf{Acknowledgments}.  We  acknowledge financial support of this research from the Air Force Office of Scientific Research under contract number FA9550-10-1-0260 and the Natural Sciences and Engineering Research Council of Canada.  Useful discussions with Prof. G. D. Scholes during the course of this work are also gratefully acknowledged.

\pagebreak


\begin{thebibliography}{100}
\bibitem{sound} A partition is ``physically sound" if, given two points, $x_i$ and $x_j$, the interaction energy between
two electrons is the same whether we assume them as belonging to an atom or to a molecule \cite{li}
\end{thebibliography}
\end{document}